\documentclass[10pt,journal]{IEEEtran}

\selectfont

\usepackage[utf8]{inputenc}
\usepackage{multirow}
\usepackage{acronym}
\usepackage{amssymb}
\usepackage{pifont}
\usepackage{rotating}
\usepackage{url}
\usepackage{soul}
\usepackage{array}
\usepackage{amsmath}
\usepackage{makecell}
\usepackage{breakurl}
\usepackage{balance}
\usepackage{xcolor}
\usepackage{hyperref}
\usepackage{xr}
\usepackage[nocompress]{cite}
\usepackage{verbatim}

\acrodef{RF}{Radio Frequency}
\acrodef{TFA}{Two Factor Authentication}
\acrodef{OTP}{One Time Password}
\acrodef{NFC}{Near Field Communication}
\acrodef{SNR}{Signal to Noise Ratio}
\acrodef{SVM}{Support Vector Machine}
\acrodef{FFT}{Fast Fourier Transform}
\acrodef{MITM}{Man In The Middle}
\acrodef{OFDM}{Orthogonal Frequency Division Multiplexing}
\acrodef{CA}{Certification Authority}
\acrodef{PKI}{Public Key Infrastructure}
\acrodef{DH}{Diffie-Hellman}
\acrodef{TTP}{Trusted Third Party}
\acrodef{IMD}{Implanted Medical Device}
\acrodef{RFID}{Radio Frequency Identification}
\acrodef{ICA}{Independent Component Analysis}
\acrodef{ToF}{Time of Flight}
\acrodef{ToA}{Time of Arrival}
\acrodef{DSP}{Digital Signal Processing}
\acrodef{FMCW}{Frequency-Modulated Carrier Waves}
\acrodef{BSS}{Blind Source Separation}
\acrodef{CAPTCHA}{Completely Automated Public Turing tests to tell Computers and Humans Apart}
\acrodef{RSH}{Robust Sound Hash}
\acrodef{DOA}{Direction Of Arrival}
\acrodef{MFCC}{Mel-Frequency Cepstral Coefficient}
\acrodef{IoT}{Internet of Things}
\acrodef{LoS}{Line of Sight}
\acrodef{ASSS}{Audio Secret Sharing Scheme}
\acrodef{RSSI}{Received Signal Strength Indicator}
\acrodef{HE}{Homomorphic Encryption}
\acrodef{OOB}{Out Of Band}
\acrodef{DoA}{Direction of Arrival}
\acrodef{ML}{Machine Learning}
\acrodef{VCS}{Voice Controllable System}
\acrodef{VA}{Voice Assistant}
\acrodef{NLP}{Natural Language Processing}
\acrodef{VUI}{Voice User Interface}
\acrodef{VOIP}{Voice Over IP}
\acrodef{EER}{Equal Error Rate}
\acrodef{MFA}{Multi-Factor Authentication}
\acrodef{MCC}{Matthews Correlation Coefficient}
\acrodef{FP}{False Positive}
\acrodef{TN}{True Negative}
\acrodef{DoS}{Denial of Service}
\acrodef{FHSS}{Frequency Hopping Spread Spectrum}
\acrodef{DSSS}{Direct Sequence Spread Spectrum}
\acrodef{KNN}{K-Nearest Neighbors}
\acrodef{MCC}{Matthews' Correlation Coefficient}

\newcolumntype{P}[1]{>{\centering\arraybackslash}p{#1}}
\newcommand{\cmark}{\ding{51}}%
\newcommand{\xmark}{\ding{55}}%

\newcommand\blfootnote[1]{%
  \begingroup
  \renewcommand\thefootnote{}\footnote{#1}%
  \addtocounter{footnote}{-1}%
  \endgroup
}

\begin{document}
\bstctlcite{IEEEexample:BSTcontrol}
%
\title{Short-Range Audio Channels Security: Survey of Mechanisms, Applications, and Research Challenges}

\author{Maurantonio Caprolu, Savio Sciancalepore, Roberto Di Pietro
\\Division of Information and Computing Technology (ICT) \\ College of Science and Engineering (CSE), Hamad Bin Khalifa University (HBKU) \\ Doha, Qatar \\
  mcaprolu@mail.hbku.edu.qa, ssciancalepore@hbku.edu.qa, rdipietro@hbku.edu.qa}

%

\IEEEtitleabstractindextext{%
\begin{abstract}
\blfootnote{This is a personal copy of the authors. Not for redistribution. The final version of the paper is available through the IEEExplore Digital Library, at the link: \url{https://ieeexplore.ieee.org/document/8967104}, with the DOI: \url{10.1109/COMST.2020.2969030.}} Short-range audio channels have appealing distinguishing characteristics: ease of use, low deployment costs, and easy to tune frequencies, to cite a few. 
Moreover, thanks to their seamless adaptability to the security context, many techniques and tools based on audio signals have been recently proposed. However, while the most promising solutions are turning into valuable commercial products, acoustic channels are also increasingly used to launch attacks against systems and devices, leading to security concerns that could thwart their adoption.
To provide a rigorous, scientific, security-oriented review of the field, in this paper we survey and classify methods, applications, and use-cases rooted on short-range audio channels for the provisioning of security services---including Two-Factor Authentication techniques, pairing solutions, device authorization strategies, defense methodologies, and attack schemes. 
Moreover, we also point out the strengths and weaknesses deriving from the use of short-range audio channels. 
Finally, we provide open research issues in the context of short-range audio channels security, calling for contributions from both academia and industry.
\end{abstract}

}

\maketitle
\IEEEdisplaynontitleabstractindextext

%
\IEEEpeerreviewmaketitle

\ifCLASSOPTIONcompsoc
\IEEEraisesectionheading{\section{Introduction}\label{sec:intro}}
\else

\begin{IEEEkeywords}
Audio Channel Security, Audio-based Authentication, Pairing via Audio, Audio Attacks, Audio for Cyber-Physical Systems Security.
\end{IEEEkeywords}

\section{Introduction}
\label{sec:intro}
\fi

\IEEEPARstart{S}ound, including human speech, is commonly considered as a natural and intuitive means to quickly interact with automatic devices~\cite{Belloch2019}. Indeed, ambient sounds, as well as voice commands issued towards audio-enabled devices, are often conceived as a natural, intuitive, and minimal effort approach for humans to communicate with machines, especially if compared to human-imperceptible \ac{RF} transmissions and distracting visual-tactile interfaces~\cite{Pan2017}.

Recently, communications using short-range audio channels have attracted increasing interest, in academia as well as industry, as demonstrated by the expected  31.80 USD billions for this specific market expected by the year 2023~\cite{Markets_2019}. 
In this context, security-oriented applications of the short-range audio stem in a prominent position.
Besides aspects related to the enhanced usability, there are mainly two dominating research directions. For the defense of systems and devices, the general opinion in the research area is that acoustic channels are a more secure communication channel when compared to legacy RF communications. For instance, given that audio channels can be perceived by participating entities, they are usually assumed to be robust against active attacks, since any malicious signal could be heard by the participants as well---and hence the attack could be detected~\cite{Hu_TDSC2018}. 
At the same time, the increasing diffusion and affordability of \ac{ML} techniques have boosted the efficiency of audio signals classification and identification, leading to low-cost and disrupting attacks. 
Indeed, recent efforts demonstrated that the sound emitted from specific devices, such as keyboards and 3D printers, leaks unique information about the specific performed task, e.g., the pressed button and the printed object features, respectively~\cite{Backes2010},~\cite{Anand2018_CODASPY}. Considering that these devices do not provide any inherent defense strategy to protect information leaked on the acoustic channel, such attacks are stealthy and potentially disrupting~\cite{Petracca2015}.

Inspired by these research challenges and by the relevant contributions in the area over the last years, in this paper we survey the mechanisms, applications, use-cases and research challenges involving the use of short-range audio channels for the provision of security services. 
We show how the peculiar features of the short-range acoustic signals, including physical proximity between communicating entities and audibility of the communications, are used to provide additional authentication schemes, pairing, context-based authorization, and defense tools, as well as to launch attacks against systems and devices. As a novel contribution, we divide and classify the scientific literature according to the provided security feature. In addition, for each of the categories identified in our survey, we explain and compare the most important contributions in the actual literature. Furthermore, we also identify some crucial research challenges, whose further investigation could unleash the full potential of audio-based security solutions, and pave the way to their large-scale adoption.

\begin{figure*}[htbp]
    \centering
    \includegraphics[width=1.8\columnwidth]{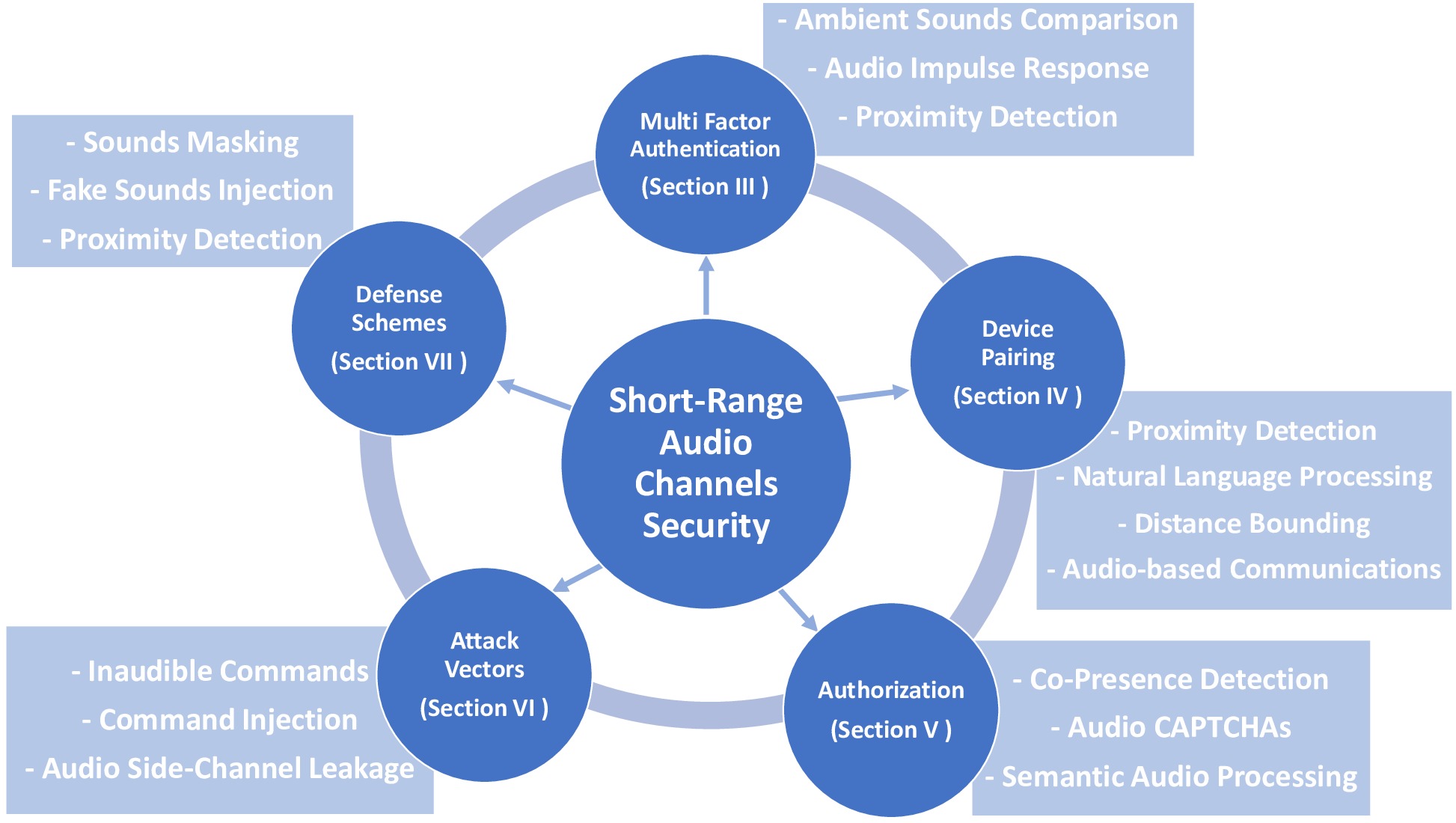}
    \caption{\textcolor{black}{ Overview of the taxonomy and the organization of our paper. We classify security schemes leveraging the audio channel based on the specific security services they intend to provide,  identifying two-factor authentication approaches (Section \ref{sec:tfa}), secure device pairing strategies (Section \ref{sec:pairing}), authorization techniques (Section \ref{sec:discrimination}), attack vectors (Section \ref{sec:attacks}), and systems defense schemes (Section \ref{sec:defense}).}
    }
    \label{fig:taxonomy}
\end{figure*}

\textcolor{black}{
Overall, the following key contributions are provided in this manuscript (see Fig. \ref{fig:taxonomy} for a graphical overview).
\begin{itemize}
    \item We survey the mechanisms, applications, use-cases and research challenges involving the use of short-range audio channels for the provision of security services to systems and users.
    \item We classify the security services provided via short-range audio channels into five (5) different categories, including authentication schemes (Section~\ref{sec:tfa}), pairing (Section~\ref{sec:pairing}), context-based authorization (Section~\ref{sec:discrimination}), attacks (Section~\ref{sec:attacks}), and defense schemes (Section~\ref{sec:defense}), detailing and cross-comparing them along the most important features.
    \item We identify several important research challenges in each of the aforementioned domains, where further scientific contributions are still required to fill the actual gaps \textcolor{black}{(Section~\ref{sec:challenges}).}
    \item \textcolor{black}{We identify a few promising future research directions involving the usage of short-range audio channels, whose thorough investigation could further unleash the potential of these communication technologies (Section~\ref{sec:challenges}).}
\end{itemize}}

\begin{table}[htbp]
\label{tab:related}
\centering
\begin{tabular}{|P{1cm}|P{7.cm}|}
\hline
\textcolor{black}{ \textbf{Paper}} & \textcolor{black}{ \textbf{Main Topic Addressed}} \\
\hline
\textcolor{black}{ \cite{Hu2018tangible}} & \textcolor{black}{ Survey of the approaches supporting the pairing between devices leveraging the physical context shared by participating entities.} \\
\hline
\textcolor{black}{ \cite{conti2018survey}} & \textcolor{black}{ Survey of strategies used to establish the co-presence between communicating devices.}\\
\hline
\textcolor{black}{ \cite{Nguyen2011authentication}} &\textcolor{black}{ Survey of multi-factor authentication techniques.}\\
\hline
\textcolor{black}{ \cite{Deepa2013}} & \textcolor{black}{Survey of attacks exploiting side-channels and unintentional information leakage.}\\
\hline
\textcolor{black}{ \cite{Kulkarni2018}} & \textcolor{black}{Survey of audio and visual CAPTCHAs used to distinguish humans from bots.}\\
\hline
\textcolor{black}{ \cite{Jayaram2011}} & \textcolor{black}{Survey of steganography and other information hiding techniques.}\\
\hline
\end{tabular}
\hfill
\caption{\textcolor{black}{Overview of related surveys and their respective topics}}
\end{table}

\textcolor{black}{
As highlighted in Tab.~\ref{tab:related}, some security solutions based on audio channels have been quickly touched in the recent literature by few surveys, focusing either on context-based security~\cite{Hu2018tangible}, ~\cite{conti2018survey}, authentication solutions~\cite{Nguyen2011authentication}, side-channel attacks~\cite{Deepa2013}, CAPTCHAs~\cite{Kulkarni2018} or forensic applications~\cite{Jayaram2011}. However, to the best of our knowledge, our contribution is the first to specifically 
survey, classify, and systematize security methods, applications, and use-cases using only short-range audio channels, as well as to evaluate such strategies across different use-cases and objectives, and to propose further research challenges---specific to this domain.}

Being focused on short-range audio channels, our survey does not include underwater communication schemes, audio watermarking, and audio forensic techniques, as they involve either long distances or considerations related to the audio signal processing, not specifically meant for secure communication and applications. 

We believe that this work could attract researchers coming from heterogeneous backgrounds, either interested in finding and applying unconventional security solutions to classical security issues, or researchers particularly skilled in the specific audio domain, finding in security applications the perfect area where they could provide different perspectives and innovative solutions. 
Indeed, through the lenses of scientifically rigorous work, the interested readers will find meaningful mechanisms and solutions related to the physical properties of audio channels used for security applications, and how they are integrated into modern computer systems and devices to enhance their security level. Further, open research challenges are also highlighted, to inspire further research in this exciting domain.

The rest of this paper is organized as follows: Section \ref{sec:background} provides the necessary preliminary details about the audio processing, as well as the most important techniques used in the literature for security services via audio. Section \ref{sec:tfa} focuses on applications of audio to Two-Factor Authentication protocols, Section \ref{sec:pairing} provides a comprehensive overview of the techniques based on audio used for the pairing of previously unknown devices, while Section \ref{sec:discrimination} details how audio can be used to authorize the usage of systems and devices. The most important attacks exploiting audio channels and their components are discussed in Section \ref{sec:attacks}, while the use of audio as a defense tool is discussed in Section \ref{sec:defense}. 
The most important research challenges arising from the above discussions are summarized and explained in Section \ref{sec:challenges}, while Section \ref{sec:conclusions} closes the paper.

\section{Background}
\label{sec:background}

In this section, we briefly introduce some technical concepts related to the analysis and usage of short-range audio signals. While a thorough analysis of these concepts is out of the scope of this contribution, their high-level description will be useful to fully catch differences and similarities between scientific contributions applying audio signals for different security-oriented tasks. Section \ref{sec:octave} illustrates the basic features of audio signals and their processing on modern computers, while the fundamental phenomena behind human speech generation and electronic speech reproduction are presented in Section \ref{subsec:human_sounds} and Section \ref{subsec:electro_sounds}, respectively. Section \ref{subsec:audio_comparison} provides an overview of the most used techniques to compare audio signals is provided, while the logic of the masking sound technique and the {\em cocktail party problem} are provided in Section \ref{sec:masking} and Section \ref{subsec:cpp}. Finally, Section \ref{sec:distance_bounding} highlights how short-range audio links can be used to estimate the distance between communicating entities.

\subsection{Analyzing Audio Signals}
\label{sec:octave}

The audible frequencies range, i.e., the range of audio signals that could be heard by human ears, spans the bandwidth [20 - 20,000] Hz. In practice, these limits fluctuate a few Hz above or below according to the particular individual, as specific features of the human ear can slightly increase or decrease sounds perception. The sounds below the minimum audible frequency are defined infra-sounds, while sounds above the 20,000 Hz threshold are ultrasounds. 

A common representation for the spectral power density in such a wide bandwidth is based on the use of octave bands, splitting the above range into 11 non-overlapping sub-bands. Analyzing audio signals in octave bands (or in their sub-components) allows for a more precise investigation of the features of the complex signal, as well as an in-depth analysis of the \ac{SNR} across the audible range~\cite{Gold2011}. 

\begin{figure}[htbp]
    \centering
    \includegraphics[width=\columnwidth]{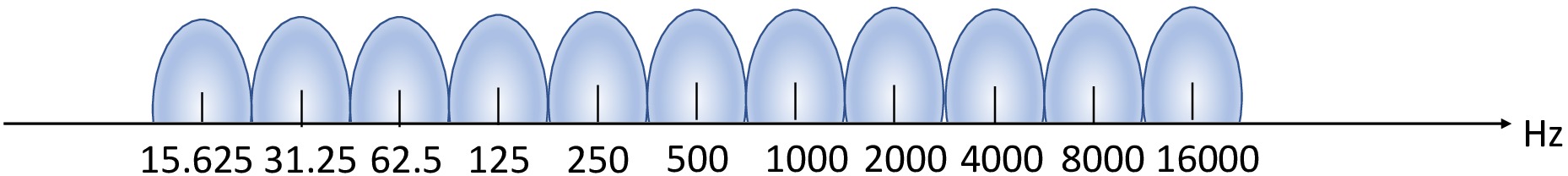}
    \caption{\textcolor{black}{ Octave bands used for audio signal representation and analysis in digital systems. There are 11 octave bands, and the center of any octave band is chosen such that it has a frequency that is exactly the double of the center frequency of the previous octave band.}}
    \label{fig:octave}
\end{figure}

\textcolor{black}{
Specifically, octave bands are selected such that the center frequency of a given band is twice the center frequency of the immediately lower band. Fig. \ref{fig:octave} shows the details of the octave bands, reported according to their calculated center frequencies.}

Note that the lower and upper limits of the audible range are essentially defined by the octave bands containing the lower and the higher frequencies that the human ear can receive.

Octave bands are mainly used when first analyzing a sound, to identify the portion of the audio spectrum where the audio signal is mainly concentrated.

In addition, octave bands are usually further divided into three ranges, namely \emph{one-third-octaves} bands. 
The use of one-third-octave bands has been standardized as a reference baseline for use in commodity scientific instruments and measurements. They are mainly used in environmental and noise control applications, to provide a further in-depth outlook on noise levels across the frequencies.

\subsection{Human Speech sounds}
\label{subsec:human_sounds}

\begin{figure}[htbp]
    \centering
    \includegraphics[width=.7\columnwidth]{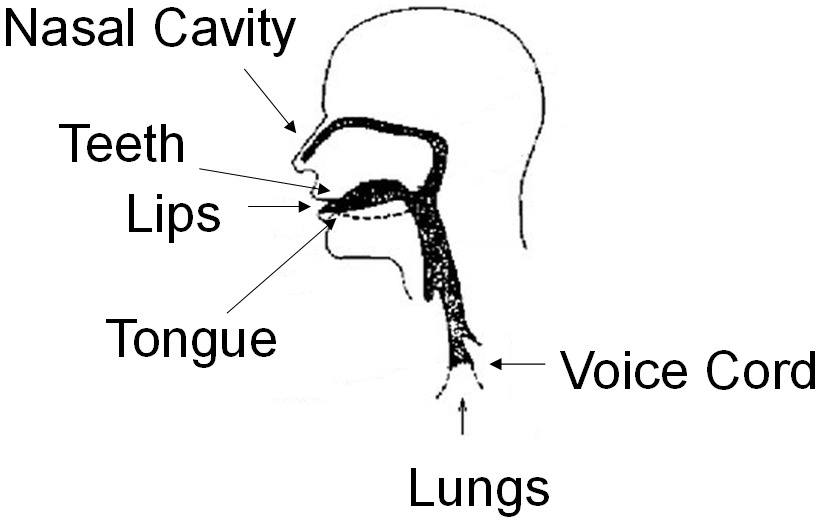}
    \caption{\textcolor{black}{ Physical elements of the human phonetic apparatus involved in speech generation. The vocal cords generate the sound, characterized by a minimum frequency, i.e., the fundamental frequency, and multiple harmonics at higher frequencies. The sound is shaped by the tongue, the lips, the teeth, and the nasal cavity, that filter out some harmonics and produce the final sound emitted by the lips.}}
    \label{fig:human}
\end{figure}

The human speech is the result of complex interactions between different elements of the human phonetic apparatus, including the vocal cords, teeth, lips, and mouth structure, to name a few, as shown in Fig. \ref{fig:human}.

\textcolor{black}{
Without loss of generality, a simplistic model is frequently used, where the speech sound is modeled as the combination of two different effects. A voiced source, physically coincident with the vibration of the vocal cords, generates the signal, characterized by a minor frequency, namely the \emph{Fundamental Frequency}, and harmonic frequencies at multiples of the fundamental one. This signal is further shaped by a filter, modeling the combined effect of the tongue, lips, teeth and nasal cavity. Overall, these components remove higher-frequencies harmonics and produce the final signal coming out from the lips. Typically, women and men have fundamental frequencies in the range [165-255] Hz and [85-180] Hz, respectively~\cite{titze1998principles}. This is caused by the structure of the vocal cords, that men have typically larger than women. However, given that no new frequencies are introduced by the elements in the mouth, the voice of an individual is typically uniquely associated with its fundamental frequency, i.e., the lowest frequency in the voice signal, also known as the \emph{pitch} of the speaker's voice. It is worth noting that the final sound produced by a human is given by the interaction between the action of the lips and the nose, further shaping the tone of the voice.}

A technique that is frequently used for speech recognition is the \ac{MFCC}, described in~\cite{muda2010voice}. This technique best describes how the human phonetic system perceives sounds. The MFCC technique divides the whole sampled window in very short windows, where the audio signal can be safely assumed as stationary. Then, the \ac{FFT} of each window is computed and it is shifted to an alternative frequency axis, namely the non-linear Mel scale, further divided into a set of bands, known as the Mel bands. This step is useful given that the speech spectrum is very wide and it does not follow a linear scale. Thus, to represent the speech signal, an alternative logarithmic scale is used, summarizing the spectrum in an increasingly wide bandwidth.   

For each Mel band, the Discrete Cosine Transform (DCT) is applied to the logarithm of the power spectral density of the input signal, to provide the final MFCC representation. There are studies, such as~\cite{hasan2004speaker}, that demonstrated the suitability of the MFCC representation for speech recognition, as it models particularly well the variability of the human speech across different users. In addition, the \ac{MFCC} technique enables further applications, such as the speaker emotion recognition~\cite{sato2007emotion}.

\subsection{Electronic Speakers Sounds}
\label{subsec:electro_sounds}

\begin{figure}[htbp]
    \centering
    \includegraphics[width=0.6\columnwidth]{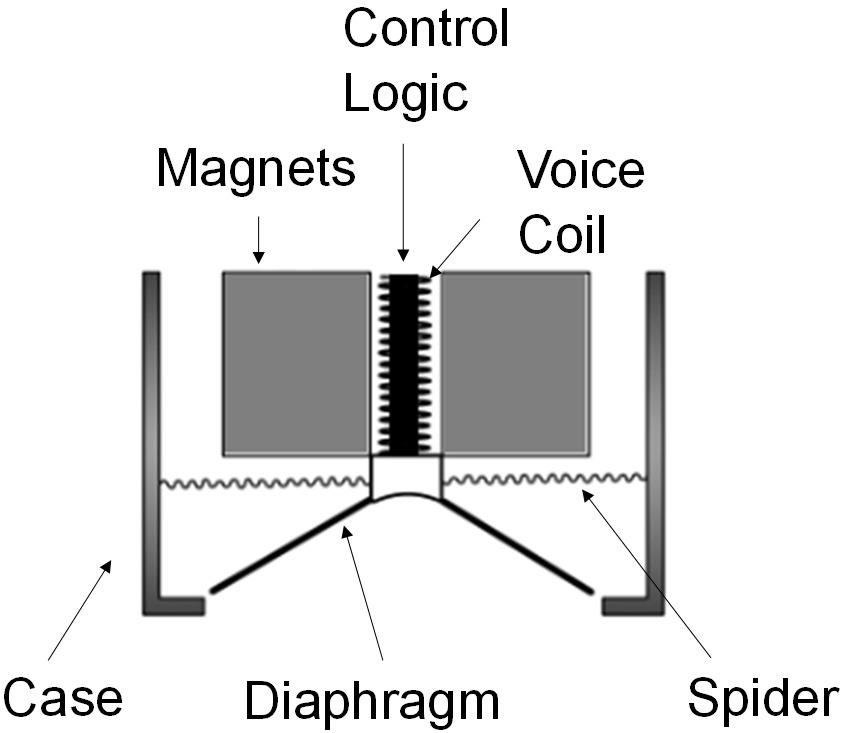}
    \caption{\textcolor{black}{ Internal components of a modern speaker. The control logic provides current to the voice coil, generating a magnetic field. The magnetic field induces the movements of the diaphragm, which modulates air pressure and produces the sound waves.}
    }
    \label{fig:speaker}
\end{figure}

The simplified internal structure of a modern sound-generation speaker is depicted in Fig. \ref{fig:speaker}.

Modern speakers reproduce sounds through the controlled oscillation of the diaphragm, steered by a control circuit and a voice coil. The control circuit receives an electronic signal from the device where the speaker is connected. The control circuit provides current to the voice coil, which further generates a magnetic field. The voice coil is located in-between two magnets, and the interaction among them causes a force that induces the diaphragm to move. The speaker also contains a spider, physically connecting the diaphragm to the case. Its role is mainly structural, as it should ensure the persistent connection between the diaphragm and the case while also allowing a free oscillation of the diaphragm and the voice coil to return in its original position. Finally, it is worth noting that the material the case consists of is crucial to not deteriorate the produced sound too much. Indeed, any material has its resonance frequency, causing resonating effects in the sub-bass region, in the frequency band [20-80] Hz.

The physical features of the speakers directly impact the performance of the device. Typically, this is expressed via the Frequency Response Curve, indicating the quality of the sounds reproduced by the speaker at a given frequency. Usually, to guarantee optimal performances, speakers are designed to work well with either low frequencies or higher frequencies, due to physical trade-offs (size of the materials). Indeed, to induce higher frequency sounds, the diaphragm should move more air, and hence, the voice coil would need to induce a stronger magnetic field by maintaining the same form factor of the speaker. Thus, it could be possible only if the voice coil is larger, i.e., by changing the physical components in the speaker. This is the reason at the basis of the choice operated by modern sound systems, that are equipped with distinct speakers, each optimized for the bandwidth of the sound to be reproduced~\cite{Blue2018_wisec}.

\subsection{Comparing Audio Recordings}
\label{subsec:audio_comparison}

Comparing audio recordings to establish their similarity is a classical problem in audio processing, very frequently faced by the approaches available in the literature. Hence, various techniques have been proposed.

A common metric to assess the similarity between the two sound registrations is established via a \emph{similarity score}. It is computed as the average of the maximum normalized cross-correlation over the pair of signal components acquired by the two devices. This metric is particularly useful when the two audio recordings are quasi-synchronized, and an indication of their similarity can be extracted from their time-domain representation.
Defining the two signals $x$ and $y$ as n-points discrete time series, namely $x(n)$ and $y(n)$, the cross-correlation $c_{x,y} (l)$ is obtained with reference to a time lag $l \in \{ 0,1 \}$, as depicted in the following Eq. \ref{eq:cross-corr}.
\begin{equation}
    \label{eq:cross-corr}
    c_{x,y} (l) = \sum_{i=0}^{n-1} x(i) \cdot y(i-l),
\end{equation}
with $y(i)=0$ when $i<0$ or $i>n-1$. Usually, two audio recordings are characterized by different amplitudes, as the receiving microphones are located at different distances from the emitting sources. To eliminate dependency from the amplitudes, a normalized measure is taken as in the following Eq. \ref{eq:norm}, to map the correlation values in the interval [-1,1]. 
\begin{equation}
    \label{eq:norm}
    c'_{x,y} = \frac{ c_{x,y}(l) }{ \sqrt{c_{x,x}(0) \cdot c_{y,y}(0) } }.  
\end{equation}
Indeed, a value $c'_{x,y}=1$ indicates that the two recordings have equal evolution in the time domain, independently from their amplitudes. A value $c'_{x,y}=-1$ indicates that they have equal evolution in the time domain but opposite signs, while the value $c'_{x,y}=0$ indicates that they are not correlated each other.

For an n-octave band's audio signal, the similarity score is finally defined as in Eq. \ref{eq:sim_score}.
\begin{equation}
    \label{eq:sim_score}
    S_{x,y} = \frac{1}{n} \sum_{i=1}^n c'_{x_i,y_i}(l).
\end{equation}

When the frequency domain features are important as well as time-domain ones, standard classification techniques can be applied.

In the following, we briefly discuss Machine Learning Techniques used for classification, Statistical Classification Methods, and finally, the Robust Sound Hash (RSH) technique.

\subsubsection{\textcolor{black}{Machine Learning Techniques}}\hfill\\
\label{subsec:ml}
\textcolor{black}{
We briefly describe the most important machine learning techniques used in the scientific contributions included in this survey, i.e., Support Vector Machines (SVM), Random Forest, and k-Nearest Neighbors (kNN), as well as the widespread k-Fold Cross-Validation technique. For interested readers, we recall that a comprehensive study of the most useful machine learning techniques for audio analysis is available in~\cite{camastra2015machine}.\\ 
\textbf{\acp{SVM}}~\cite{Cortes1995} represent a supervised learning model for classifying data, i.e., assessing which pre-defined class is more similar compared to a new set of data. As a supervised learning technique, starting from a set of \emph{training data} labeled on purpose as belonging to one or more categories, a \ac{SVM} can classify new data as being part of a class or another. In addition, a set of unlabelled data, namely \emph{testing data}, are provided for classification. Assuming $s$ categories, the \ac{SVM} internally builds a $s-1$-dimensional hyperplane in an $s$-dimensional space leveraging the labeled data, and maps the unlabelled one in this space. The output classification is provided as the category whose training points are the nearest to the testing data.\\
\textbf{Random Forest}~\cite{breiman2001random} is an ensemble supervised machine learning technique, built as a combination of decision tree predictors. As an ensemble learning technique, the provided classification is the result of a decision taken collectively, from a large number of classifiers. The idea behind ensembles classification is based upon the premise that a set of classifiers can provide a more accurate and generalized classification than a single classifier, thus being less prone to over-fitting. When using Random Forest classifiers, each classifier is a tree, and each tree depends on the values of a random vector independently sampled, assuming the same distribution for all the trees in the forest.\\
\textbf{\ac{KNN}}~\cite{aha1991instance} is a supervised machine learning algorithm able to solve both classification and regression problems. This methodology is based on the observation that instances with similar properties are generally found close together within a dataset. The classification method determines the label of an unclassified object, by observing the classes of its nearest neighbors. Thus, the algorithm determines the single most frequent class label in the set of labels of the $k$ instances nearest to the unclassified object.\\
\textbf{K-Fold Cross-Validation}~\cite{kohavi1995study} is an accuracy estimation method that allows evaluating how the results of a model can be generalized to an independent dataset. The main objective of cross-validation methods is to estimate the generalization of a model, i.e., to understand its accuracy in the classification of data that it had never seen before, minimizing the effect of overfitting problems. The method consists in partitioning the dataset into subsets, some of them (the training set) used to perform the training of the model, and the remaining ones to be used for validation (the validation set) or testing (the testing set) purposes.
Overall, it is worth noting that the \ac{SVM} technique leads to better performances when dealing with multiple dimensions and continuous features, achieving its maximum prediction accuracy when the sample size is large. The Random Forest technique, instead, performs better when dealing with numerical and categorical features. Random Forest returns the probability of belonging to a class, while SVM calculates the distance to the boundary. Without loss of generality, we notice that no single techniques can outperform other algorithms over all datasets. Thus, the main issue when dealing with classification problems is to understand the conditions where a particular technique can significantly outperform other algorithms on a given application~\cite{kotsiantis2007supervised}.}\\

\subsubsection{Other Statistical classification Metrics}\hfill\\
\label{subsec:classification}

In the context of audio signal classification, the typical features used for optimal audio classification are the time and frequency correlation. While the time-domain correlation features the same logic previously described, the frequency domain correlation can be expressed via the Pearson's correlation coefficient. Defining $X(k)$ and $Y(k)$ the \ac{FFT} of the two audio recordings $x(t)$ and $y(t)$, with $k=1...m$, respectively, the Pearson's correlation coefficient is defined in Eq. \ref{eq:pearson}, as: 
\begin{equation}
    \label{eq:pearson}
    s_f(X,Y) = \frac{ \sum_{k=1}^m \left( X(k) - \bar{X} \right) \left( Y(k) - \bar{Y} \right) }{\sqrt{ \sum_{k=1}^m  \left( X(k) - \bar{X} \right)^2 \sum_{k=1}^m \left( Y(k) - \bar{Y} \right)^2  }},
\end{equation}
where $\bar{X}$ and $\bar{Y}$ are the mean of the vectors $X(k)$ and $Y(k)$, respectively~\cite{Benesty2009}. 

\textcolor{black}{
Another coefficient, widely used to decide whether a prediction is correlated to the data or it is closer to a random guess, is the \ac{MCC}, defined as in Eq.~\ref{eq:matthews} ~\cite{MATTHEWS1975442}.
\begin{equation}
    \label{eq:matthews}
    |\ac{MCC}| = \sqrt{\frac{\chi^2}{n}},
\end{equation}
where $\chi$ is the chi-square~\cite{lancaster2005chi} of the variables representing the two audio recordings.\\
Some other metrics can be used in descriptive statistics, such as the largest and the smallest observation values, defined as the values of the greatest and least elements of a sample, respectively.\\
A few classification systems used time-domain features in combination with other ones, since alone they are often not enough descriptive~\cite{gerhard2003audio}. An example is the classification system presented by the authors in~\cite{scheirer1997construction}, which uses the duration of harmonic segments as a feature to identify human speech on audio samples. \\  
Table~\ref{tab:class-comparison} provides a summary of the most used correlation/similarity classification techniques in the scientific contributions included in this survey.}

\begin{table}[htbp]
\centering
\begin{tabular}{|c|c|c|c|lllll}
\cline{1-4}
{ \textbf{\begin{tabular}[c]{@{}c@{}}\textcolor{black}{Classification}\\ \textcolor{black}{Technique}\end{tabular}}} & { \textbf{\textcolor{black}{Description}}} & { \textbf{\begin{tabular}[c]{@{}c@{}}\textcolor{black}{Time} \\ \textcolor{black}{Domain}\end{tabular}}} & { \textbf{\begin{tabular}[c]{@{}c@{}}\textcolor{black}{Freq.} \\ \textcolor{black}{Domain}\end{tabular}}} &  &  &  &  &  \\ \cline{1-4}
{ \begin{tabular}[c]{@{}c@{}}\textcolor{black}{Pearson's} \\ \textcolor{black}{Correlation} \\ \textcolor{black}{Coefficient}\end{tabular}} & { \begin{tabular}[c]{@{}c@{}}\textcolor{black}{Expresses the frequency}\\ \textcolor{black}{domain correlation.}\end{tabular}} & { \textcolor{black}{\xmark}} & { \textcolor{black}{\cmark}} &  &  &  &  &  \\ \cline{1-4}
{ \textcolor{black}{MCC}} & { \begin{tabular}[c]{@{}c@{}}\textcolor{black}{Check if prediction is} \\ \textcolor{black}{correlated to the data.}\end{tabular}} & { \textcolor{black}{\xmark}} & { \textcolor{black}{\cmark}} &  &  &  &  &  \\ \cline{1-4}
{ \begin{tabular}[c]{@{}c@{}}\textcolor{black}{Similarity}  \\ \textcolor{black}{score}\end{tabular}} & { \begin{tabular}[c]{@{}c@{}}\textcolor{black}{Evaluates the similarity} \\ \textcolor{black}{between two sounds.}\end{tabular}} & { \textcolor{black}{\cmark}} & { \textcolor{black}{\xmark}} &  &  &  &  &  \\ \cline{1-4}
{ \textcolor{black}{MFCC}} & { \begin{tabular}[c]{@{}c@{}}\textcolor{black}{Represents the short-term} \\ \textcolor{black}{power spectrum} \\ \textcolor{black}{of a sound.}\end{tabular}} & { \textcolor{black}{\xmark}} & { \textcolor{black}{\cmark}} &  &  &  &  &  \\ \cline{1-4}
{ \textcolor{black}{Peak Ratios}} & { \begin{tabular}[c]{@{}c@{}}\textcolor{black}{Considers the peak values} \\ \textcolor{black}{of two variables.}\end{tabular}} & { \textcolor{black}{\cmark}} & { \textcolor{black}{\cmark}} &  &  &  &  &  \\ \cline{1-4}
{ \begin{tabular}[c]{@{}c@{}}\textcolor{black}{Specific Band}\\ \textcolor{black}{Energy}\end{tabular}} & { \begin{tabular}[c]{@{}c@{}}\textcolor{black}{Considers the energy value}\\ \textcolor{black}{in a specific octave band.}\end{tabular}} & { \textcolor{black}{\xmark}} & { \textcolor{black}{\cmark}} &  &  &  &  &  \\ \cline{1-4}
\end{tabular}
\vspace{1.0pt}
\caption{\textcolor{black}{Most popular correlation/similarity classification techniques.}}
\label{tab:class-comparison}
\end{table}

\subsubsection{\acl{RSH}}\hfill\\
Another technique used to compare audio recordings is the \acl{RSH} (RSH), especially when the exact matching between the audio recordings is desired. Ideally, to compare two audio recordings, a viable solution would be to compute the digest of each audio recording samples via a standard cryptographic hashing function, and to verify the matching of the two digests. However, as per the basic properties of cryptographic hashing functions, even small differences in source files would lead to totally different digests~\cite{stallings2017cryptography}. Thus, in the context of audio files, where the background random noise is always present, they are not applicable. A valuable alternative is the use of comparison mechanisms that change slightly in response to a minor variance of the input. An example is the \ac{RSH} technique, designed in~\cite{jiao2009robust}. The RSH divides the input file in different frames, each containing 1s of recording. Then, it applies a specific function to this digest, and it outputs a fixed-length string representing the sampling interval. The digests produced for each time frame are concatenated to form the final output. Then, the similarity between the two audio files is computed as the \emph{Hamming Distance} between the two digests, i.e., the number of dissimilar bits. Because of the unique features of the function used in signal processing, similar audio recordings will have low dissimilar values. Thus, a threshold can be established to determine if two hashes are similar enough to be considered as matching recordings.

\subsection{Masking Sounds}
\label{sec:masking}

\begin{figure}[htbp]
    \centering
    \includegraphics[width=\columnwidth]{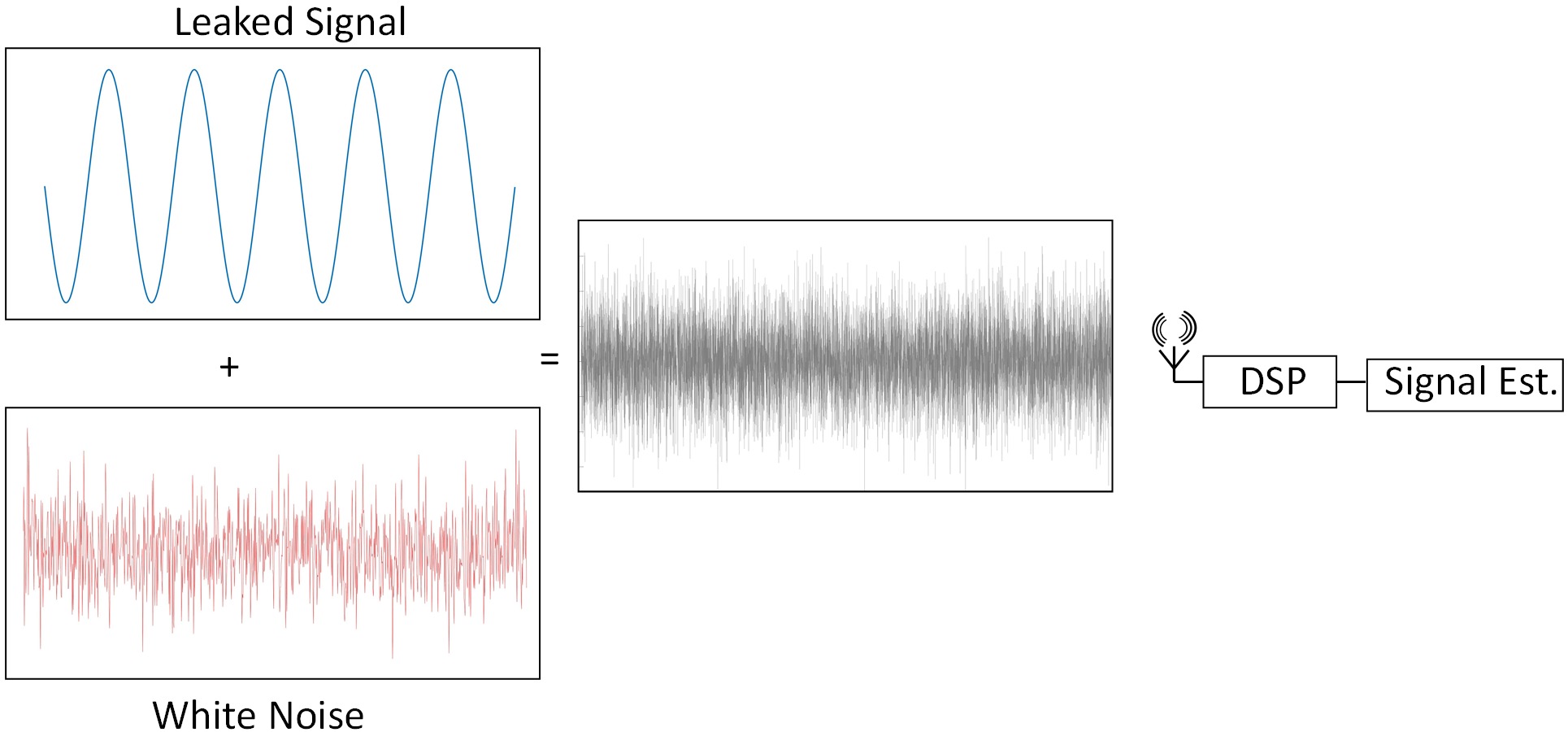}
    \caption{\textcolor{black}{ Logic of sound masking mechanism. White noise produced on purpose sums up to the signal leaked from a given device, in order to corrupt it and make it unrecoverable to an adversarial Digital Signal Processing logic.}
    }
    \label{fig:audio_defense}
\end{figure}

The \emph{Sound Masking} technique is frequently used in audio-based defense systems to protect against unauthorized eavesdropping. It inherits its core logic from the \emph{Friendly Jamming} technique, widely used in the wireless security research domain to hide sensitive information or block any other communication other than the authorized ones~\cite{Vilela2010},~\cite{Shen2013}.
The main logic of the sound masking strategy is illustrated in Fig. \ref{fig:audio_defense}.

The audio leakage from a system, e.g. a keyboard or a vibration-based communication system, as thoroughly described in Section \ref{sec:attacks}, can lead to the identification of privacy-sensitive information, such as the pressed key or the sound's vibration. To mislead an external attacker, white noise is produced typically by an additional device coupled with the main communication channel, and it is emitted on purpose to taint the information signal. The malicious device, getting the mixed-signal, has the hard task of separating emitting components through \ac{DSP} techniques to estimate the useful signal. 

\subsection{Separating Sounds: The Cocktail Party Problem}
\label{subsec:cpp}

The audio masking strategy described in Section \ref{sec:masking} is effective only if an external third-party cannot perform a separating attack, i.e., it cannot decouple the jamming signal from the leaked information signal. Indeed, an adversary could perform such an attack in two different phases. In an online phase, it could gather the mixture signals via multiple microphones arranged randomly in the environment. Then, it could try to estimate the two addends, i.e., the information signal and the jamming signal. Assuming to have two different receivers and two different signal components $s_1$ and $s_2$, the mixtures $x_1$ and $x_2$ recovered by the microphones are expressed by Eq. \ref{eq:mix}.
\begin{equation}
    \label{eq:mix}
    x = H \cdot s + e \rightarrow \begin{bmatrix} x_1 & x_2 \end{bmatrix} = \begin{bmatrix} h_{1,1} & h_{1,2} \\ h_{2,1} & h_{2,2} \end{bmatrix} + \begin{bmatrix} e_1 & e_2 \end{bmatrix},
\end{equation}
where $e_1$ and $e_2$ represent the noise associated with each sampling of the audio signal by the receivers.

In the literature, this task is commonly referred to as the \emph{Cocktail Party Problem}---in an analogy with the task of separating the linear mixture of different voices at a cocktail party, where voices of different tones overlap each other~\cite{Haykin2005}. The class of approaches used to solve this problem is called \ac{BSS}~\cite{comon2010handbook}. There are many \ac{BSS} algorithms available in the literature, each one based on specific assumptions on the signals involved in the mixtures~\cite{pedersen2007survey}. One of the most successful is the \ac{ICA} technique, whose goal is to express the mixed signal as a linear combination of non-Gaussian components, in a way that they are as much statistically independent as possible~\cite{hyvarinen2000independent}. Regarding Eq. \ref{eq:mix}, this is equivalent to find an un-mixing matrix $W$, such as the following Eq. \ref{eq:unmix} is verified.
\begin{equation}
    \label{eq:unmix}
    \hat{s} = \begin{bmatrix} \hat{s}_1 & \hat{s}_2 \end{bmatrix} = W \cdot x  = \begin{bmatrix} w_{1,1} & w_{1,2} \\ w_{2,1} & w_{2,2} \end{bmatrix} \cdot \begin{bmatrix} x_1 & x_2 \end{bmatrix},
\end{equation}

Note that, if it is successful, \ac{ICA} can recover the original mixing sources less than a scaling factor and a different naming (ordering) of the signals. However, in the context of audio mixtures, \ac{ICA} may face applicability issues when the sources of the signals are very close to each other. In this case, the algorithm cannot precisely identify the exact number of the sources, or it confuses part of a signal with the other, resulting in degraded performance~\cite{Zhang2014}.

\subsection{Distance Bounding via Audio}
\label{sec:distance_bounding}

\begin{figure}[htbp]
    \centering
    \includegraphics[width=.6\columnwidth]{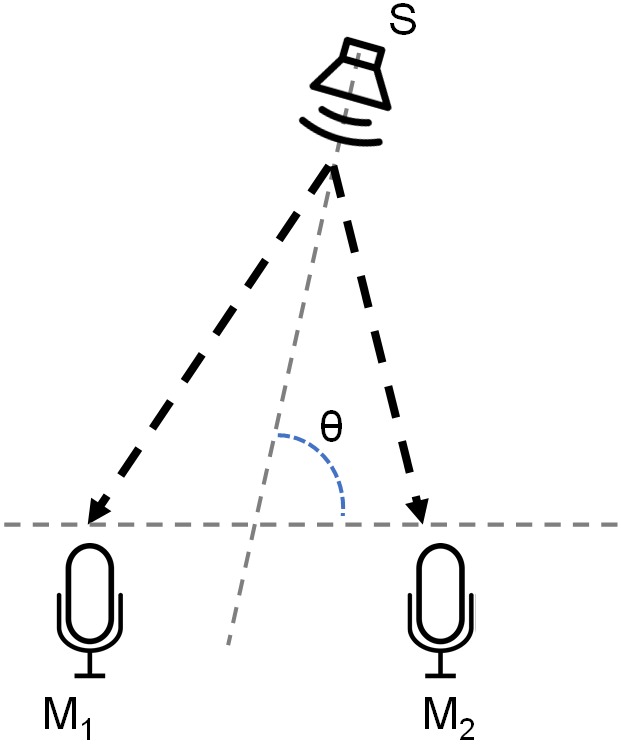}
    \caption{\textcolor{black}{ Audio-based distance bounding. At least two receiving microphones $M_1$ and $M_2$ are used to differentiate the timestamps of the signals received from a single source speaker $S$, and to estimate the distance between the transmitter and the receivers.}
    }
    \label{fig:bounding}
\end{figure}

Like any other wireless signal, short-range audio signals can be used to provide location estimations. Specifically, looking at the signal emitted by the speakers and recorded back by the microphone, the emitting device could get a rough estimate of the device-user distance, further useful for advanced security tasks. Fig. \ref{fig:bounding} illustrates a sample scenario.\\
\textcolor{black}{
The simplest way to estimate the distance between a user and a device is to recur to the timestamps of the emitted and received signals. Let us assume that a single receiver is available, namely $M_1$, that $t_0$ is the time in which the sound is emitted and $t_r$ is the reception time on the microphone $M_1$. The \ac{ToF} can be obtained as $ToF = t_r - t_0$. Defining $c$ the speed of sound, conventionally equal to 340 m/s, the distance between the user and the device can be easily obtained as in the following Eq. \ref{eq:tof}.
\begin{equation}
    \label{eq:tof}
    d = c \cdot \frac{ToF}{2}.
\end{equation}\\
While being extremely simple to implement, the results obtained through the above formula, unfortunately, are affected by large errors, especially because of the large inaccuracies in timestamps gathered on mobile devices, such as smartphones and wearables, caused by various delays introduced in the processing chain~\cite{Scott2005},~\cite{Peng2007}. Thus, two techniques are mainly used.
Where the timestamps of the signal are the only available data to be used and the emitting frequency cannot be changed, the \ac{DoA} technique is a viable option. DoA resorts to the time of arrivals of the signals on at least two receivers (in the case of audio signals, microphones), to determine the location where the source is. Assume $t_1$ and $t_2$ to be the reception timestamps of a given signal emitted by a source $S$ and received by two receivers, $M_1$ and $M_2$, respectively. Defining $\tau = t1-t2$, the angle of arrival $\theta$ can be computed as in the following equation Eq. \ref{eq:doa}.
\begin{equation}
    \label{eq:doa}
    \theta = cos^{-1} \frac{c\tau}{d}.
\end{equation}\\
Eq. \ref{eq:doa} provides a single direction over which the source is located. The process can be repeated for every couple of considered receivers, providing an equal number of directions and thus a precise estimate of the specific point where the source lies. At the same time, inaccuracies due to environmental noise or hardware imperfections could be reduced.\\
An even more effective strategy to achieve source localization, namely \ac{FMCW}, is discussed in~\cite{Mahafza2013}, with reference to radar systems, and it is based on frequency sweeps. The basic idea of this approach is to estimate the ToF indirectly, via differences in the frequency between transmitted and received signals. \ac{FMCW} is based on a given number of rounds, being the final result the median of the results provided by single rounds. Assume that $f_0$ is the starting frequency and $f_1$ is the ending frequency, while $T_s$ is the time needed by the transmitter to span the spectrum range from $f_0$ to $f_1$ in linear time at a pre-defined speed. At a given time, the transmitting device (i.e., the speaker) emits a sound on a frequency, that will be received by the receiving device on a given frequency. Defining $\Delta_f$ the frequency shift, i.e., the difference between the transmitting and receiving frequency, then the ToF, namely $\Delta t$, can be computed as in the following Eq. \ref{eq:tof_2}.
\begin{equation}
    \label{eq:tof_2}
    \Delta t = \frac{\Delta_f}{f_1-f_0} \cdot T_s.
\end{equation}\\
Note that, Eq. \ref{eq:tof_2} does not contain the reception timestamps, but the transmitting and receiving frequency. Given that the transmitting frequency is chosen by the transmitter while the receiving frequency can be precisely estimated via standard \ac{FFT}, the resulting accuracy can be highly improved. Once the ToF is obtained, the distance device-user can be obtained as in the following Eq. \ref{eq:d}.
\begin{equation}
    \label{eq:d}
    d = c \cdot \Delta t.
\end{equation}\\
While the solutions discussed above will be further discussed throughout the paper, many other techniques to achieve source localization using acoustic signals are available. We refer the interested readers to the contributions already available in the literature, surveying methods for acoustic source localization~\cite{Belloch2019},~\cite{Chiariotti2019},~\cite{Liu2018}.}

\textcolor{black}{
\section{Audio Channels for Two-Factor Authentication}
\label{sec:tfa}
In this section, we provide details on the use of audio channels as a second factor for the authentication of a user to a system or a device. The background on multi-factor authentication technique is provided in Section \ref{subsec:back_tfa}, while the literature overview and the comparisons are provided in Section \ref{subsec:tfa_literature}.}

\subsection{Background}
\label{subsec:back_tfa}
\textcolor{black}{
\acl{TFA} (TFA) techniques are becoming increasingly pervasive and diffused in the last years, driven by ever increasing password leakage events\footnote{\url{https://en.wikipedia.org/wiki/List_of_data_breaches}}~\cite{Onaolapo2016}. In addition} to traditional password-based mechanisms, \ac{TFA} techniques adds a registered token in possession of the user to further enforce authentication and verify that the user credentials have not been leaked. Nowadays, \ac{TFA} is successfully adopted in the context of online banking~\cite{Kiljan2016}, enterprise access~\cite{Theofanos2016}, and mobile social networks~\cite{akhtar2019content}, resorting to dedicated private solutions or commercial services, such as Encap Security~\cite{encap_2step}, Duo~\cite{duo_2step} and Google 2-step verification~\cite{google_2step}, to name a few. 

\begin{figure}[htbp]
    \centering
    \includegraphics[width=\columnwidth]{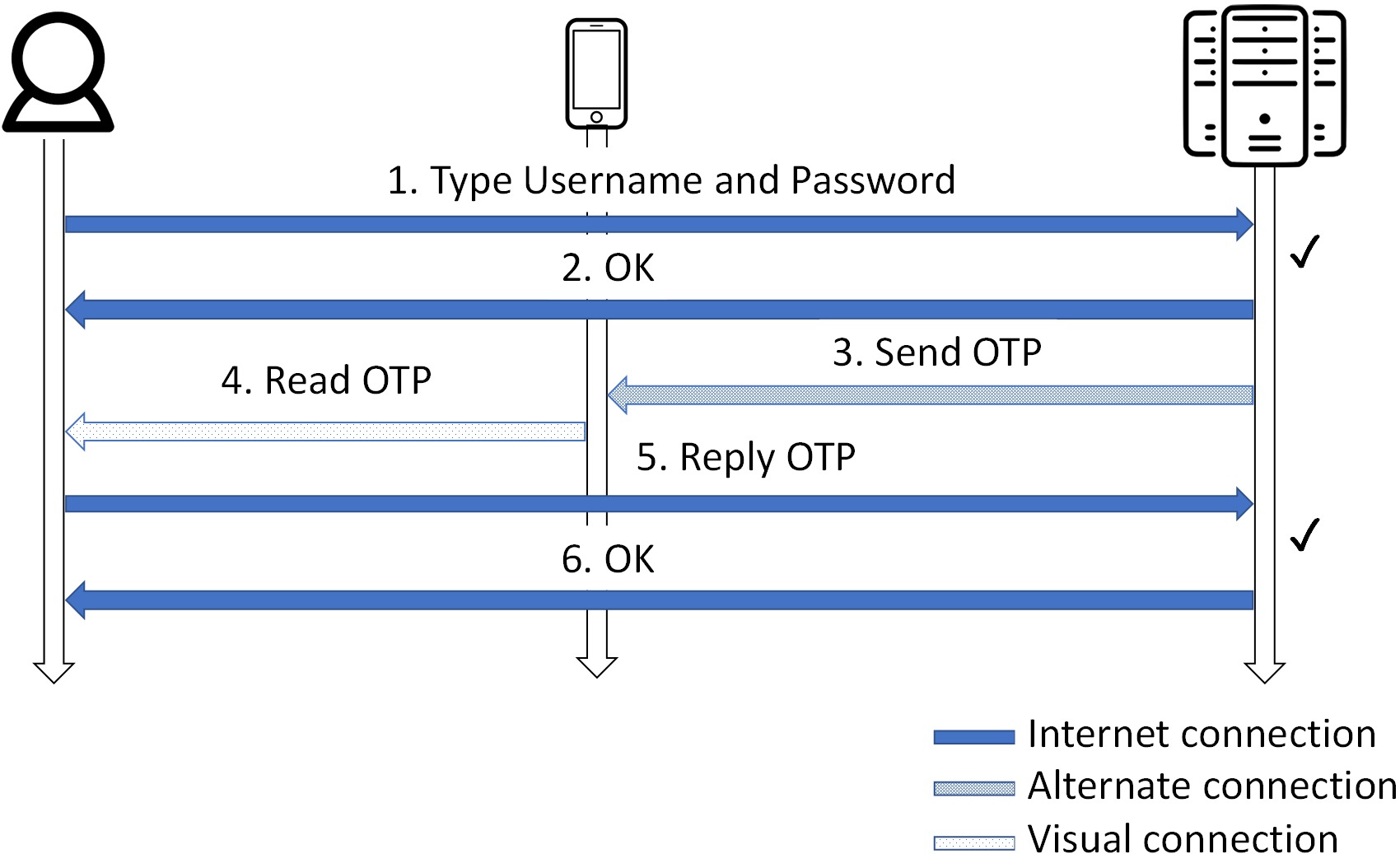}
    \caption{\textcolor{black}{ Two-Factor Authentication: example of the interactions. In the first phase, the user performs the login using its credentials. Then, to further verify its identity, the server delivers a \ac{OTP} to the mobile phone registered with the user, and it evaluates the reply provided by the user. If it matches the expected output, the physical individual behind the machine has both the credentials and the mobile phone registered by the user. Thus, it is authenticated.}
    }
    \label{fig:tfa}
\end{figure}

The most widespread form of two-factor authentication is based on the use of electronic tokens, uniquely tied to the owner, such as smartphones. In the first phase, the user inserts its credentials. After successful verification of the credentials, the server in charge of the authentication generates a \ac{OTP} and delivers it to the user. The OTP is then used, in different forms, to provide a unique response to the server and further validate the possession of a secondary device. For instance, in case the electronic token is a smartphone, the OTP is delivered via SMS to the registered telephone number and the user is requested to insert it as it is received. The diagram in Fig. \ref{fig:tfa} highlights the described interactions.

\textcolor{black}{
While the security offered by TFA techniques is undoubtedly higher than traditional mechanisms based on credentials only, its adoption, optional in most of the cases, is still not diffused. Indeed, most users still prefer the single authentication shot, mainly because of the extra effort required by legacy TFA techniques, always requiring explicit user interactions and not suitable for blind or visually impaired users~\cite{Weir2009},~\cite{Gunson2011}.\\
These drawbacks have motivated significant efforts in the last years, both by industries and academia, to develop more usable TFA schemes, eventually requiring zero interactions from the user. While some solutions based on the use of the Bluetooth~\cite{Czeskis2012} and \ac{NFC}~\cite{yubikey} technologies exist, they require the involved devices to feature a shared alternative \ac{RF} communication technology, not always available in modern devices~\cite{Velasquez2018}.\\ 
In this context, the audio communication channel can provide an effective answer. Indeed, elements such as microphones and speakers are widely accepted, and already present in several devices, ranging from regular laptops to ultimate wearables.}

\subsection{Literature Overview}
\label{subsec:tfa_literature}
\textcolor{black}{
An early but successful approach to the use of audio for TFA was provided in~\cite{Karapanos2015}, leveraging ambient sounds. The idea behind} this approach is to demonstrate that the two involved parties, i.e., the client requesting authentication and the server in charge of authenticating requests, are in each other proximity, i.e., in the same physical environment. The resulting TFA mechanism, namely \emph{Sound-Proof}, does not require any further user interaction, as the second factor for authentication is automatically enforced by triggering simultaneous registration of ambient sounds from the browser and the phone. The two-sides registrations of the ambient sounds are compared on the phone and, if significant similarities are found, the proximity of the client is established and the authentication is verified. The similarity between the two audio recordings/snippets is established via the \emph{similarity score}, as outlined in Section \ref{subsec:audio_comparison}. Thus, the two parties access the audio channel only once, for quasi-simultaneous recordings. Then, the browser delivers its audio to the smartphone, via a regular data connection, for audio file comparison. This architectural choice is mainly applied due to privacy concerns. Indeed, if the evaluation of the audio recordings takes place on the server, a potentially malicious server could have access to the ambient sounds registered by the users, leaking sensitive information. Instead, Sound-Proof moves the comparison on the smartphone, while the server is only informed about the output of such operation. \emph{Sound-Proof} only leverages ambient sounds, easily modifiable by a malicious third-party. Indeed, an adversary can deliberately make the phone create a given sound, or wait for the phone to emit a specific sound, such as a ring, a notification or an alarm sound. In these cases, the ambient audio fingerprint is indeed easily predictable, and it would allow an adversary to overcome physical barriers to provide a successful attack. This is the rationale supporting the Sound-Danger attack disclosed in~\cite{Shrestha2016}, where the authors discuss several smart attacks via induced sounds on the phones, leading to a maximum of $83.2 \%$ efficacy in compromising user accounts leveraging the Sound-Proof technique. 

An effective technique to overcome the Sound-Danger attack was provided in~\cite{Shrestha2018}, with the assistance of a smart wearable device. In this scheme, namely \emph{Listening-Watch}, the browser generates a random (i.e., unpredictable) audio signal, that is registered by the phone and the wearable device. The login succeeds if the audio recording on the wearable device contains a specific pattern and is similar enough to the recording of the browser, thus ensuring the proximity of the wearable to the browser. The similarity between the two recordings is established via the Similarity Score, as in~\cite{Karapanos2015}. The performance of \emph{Listening-Watch} have been evaluated in an office environment, with different distances between the wearable and the browser, i.e., benign (30 cm), intimate (50 cm) and personal (from 50cm to 1 m), as well as with different levels of the audio signal, i.e., full volume (100\% of the scale, 79 dBA), average volume (75\% of the volume, 74 dBA) and low volume (50\%, 67 dBA). A potential weakness of the \emph{Listening-Watch} technique is to leverage audible sounds, emitted on purpose by the speakers. This could create potential usability issues, especially when the user needs to perform the login in specific areas, such as a library or during a talk. This issue is overcome by the authors in~\cite{Wang2018}, employing ultrasounds ([18-22] kHz). The resulting technique, namely \emph{SoundAuth}, does not disturb the surrounding audio environment. In addition, it employs the \ac{SVM} technique for the audio comparison, resulting in a higher level of accuracy especially in attack situations. The security of \emph{SoundAuth} has been evaluated directly against~\cite{Karapanos2015} in different environments, including an office, a desk with underlying music, a lecture and underlying TV audio noises. Thanks to the introduction of the SVM classifier, \emph{SoundAuth} is characterized by lower levels of False Rejection Rate and False Acceptance Rate, leading to better performances. 

\textcolor{black}{
Recently, the authors in \cite{Truong2019_percom} proposed \emph{DoubleEcho}, a new context-based co-presence verification scheme that is effective in verifying that two (or more) devices are located in the same physical context. To this aim, \emph{DoubleEcho} leverages acoustic Room Impulse Response (RIR), i.e., a time-domain representation of the modifications of a particular source signal at a given distance from the source in a particular physical location. Being dependant on both sound wave propagation properties and shapes and materials from the enclosure, the RIR is very difficult to be imitated or reproduced when not in the same physical environment of the participating devices. In this scheme, the devices initiate the recordings of the audio environment only when they are triggered by one of the participating entities, and they acquire audio samples for about 2 seconds. Given that the recording captures any sound from the surrounding environment within a half meter distance, the scheme could be not fully privacy-preserving: in fact, chances are that the recordings include background information, that could lead to the identification of the environment where the devices are, or to the leakage of additional private information. The scheme has been extensively evaluated, considering different smartphones and tablet brands, and a variety of public and private environments (including offices, kitchen, corridor, classroom, meeting rooms, to name a few).
\emph{DoubleEcho} significantly improves the state-of-the-art in the mitigation of context-based context manipulation attacks, requiring a MITM attacker to be less than $50$~cm apart from the intercepted devices. 
For distances less than $50$~cm, given that successful co-located MITM attacks are still possible, the protection is delegated to the authentication of the channel between the involved devices, that can be either Bluetooth, NFC, or WiFi.\\}
\textcolor{black}{
Audible sounds are also used in the recent proposal by the authors in \cite{Choi2018}. In this contribution, the authors proposed a \ac{TFA} technique to strengthen the authentication of a wireless key-fob to a car, leveraging ambient sounds to overcome relay attacks. More in detail, at the reception of a wireless input from the remote key, the car produces random sounds in the audible frequency range, and both of the involved entities record the audio environment. The key delivers its recording to the car, where it is compared against the locally recorded sound context, using the \emph{similarity score} metric. If the two recordings are similar enough, the key is successfully authenticated; otherwise, it is rejected. The method has been evaluated both in an office and in a real car-parking scenario, with a minimum \ac{EER} of 0.0013. It is worth noting that, as for the proposal by the authors in \cite{Karapanos2015}, the reliance on audible sounds exposes the method to the Sound-Danger attack. In addition, when the adversary is co-located and can retrieve the locally stored secrets from the key-fob, the protocol is also vulnerable to \ac{MITM} attacks. Finally, being based on ambient sounds recordings, the protocol is also not privacy-preserving.}

Despite the improvement in performance and usability, the previous approaches still do not provide an answer for a co-located \ac{MITM} attack. Indeed, an adversary that is co-located with the end-device can stealthily relay messages between the two entities, bypassing the proximity check, as shown in~\cite{Drimer2007} and~\cite{Francillon2011}. A solution to this attack is provided by the \emph{Proximity-Proof} scheme, described in~\cite{Han2018}. It overcomes the co-located MITM attack through the extraction of the physical-layer fingerprints of the speaker and the microphone, thus authenticating the device involved in the communication. Thus, the TFA is approved not only if the audio environment similarity is verified, but also if the speaker and the microphone fingerprints are matched against the well-known profile stored at the server-side. The fingerprints of the microphone and the speaker are created by evaluating the effect of the transmitter and the receiver on transmission at a given frequency, spanning from 18.1 kHz to 20 kHz, as detailed in~\cite{Chen2017}. More iterations further refine the process, creating a more accurate fingerprint. The drawback of the fingerprint usually stems from the large number of audio samples required to establish the authenticity of the involved devices. Indeed, while the other approaches described above use only a single audio access for each involved device, in \emph{Proximity-Proof} the phone speaker continuously delivers the virtual packet for $m \ge 1$ times, where $m$ is a pre-defined parameter. The second-factor login is validated only if the login browser can establish the authenticity of the remote device with a probability $(1-p)^m$, $p$ being the probability that the packet can be successfully decoded. As in~\cite{Wang2018}, \emph{Proximity-Proof} leverages inaudible sounds, coupling them with the \ac{OFDM} technique and the ON-OFF modulation scheme, while a simple Euclidean Distance method is used for audio comparison.

Tab. \ref{tab:tfa} wraps up the above discussion and summarizes the most important features of the described approaches. \textcolor{black}{ It is worth noting that the performance of the described approaches are reported using either the EER, or the percentage of \acl{FP} and \acl{TN}, or the Overall Error Rate (ER).}

\begin{center}
    \small\addtolength{\tabcolsep}{5pt}
    
\begin{table*}[htbp]
\begin{tabular}{|P{0.65cm}|P{1.cm}|P{1.05cm}|P{1.1cm}|P{0.9cm}|P{1.04cm}|P{1.28cm}|P{0.9cm}|P{1.05cm}| P{1.4cm}|}
\hline
\multirow{2}{*}{\textbf{\color{black}Scheme\color{black}}}    & \textbf{Feature} & \textbf{Audio Analysis Strategy} & \textbf{Sound-Danger attack robustness} & \textbf{Privacy preserv.} & \textbf{Number of Audio   Accesses} & \textbf{Co-located MITM attacks robustness} & \multicolumn{2}{p{1.75cm}|}{\textbf{Performance Evaluation}} & \color{black} \textbf{Perf.} \\ \cline{8-9} 
                                      &                                              &                                                                                                    &                                                                                                            &                                                                                                                                                                                                       & &                                                                                                               & \textbf{Env.}       & \textbf{Dist.} &      \\ \hline

\cite{Karapanos2015} & Ambient Sound & Correlation (Sim. Score)  & \xmark & \checkmark  & 1 per device  & \color{black} \xmark & Office, Music, Lecture, TV  & Desk, Pocket &  \color{black} \mbox{$EER=0.07$}                     \\ \hline
\cite{Shrestha2018}  & Audible Sound & Correlation (Sim. Score)  & \checkmark & \xmark & 1 per device  & \color{black} \xmark & Office  & Intimate, Personal, Benign   & \color{black} \mbox{$EER<0.01$}                  \\ \hline
\cite{Wang2018}      & Ultrasound & SVM & \checkmark & \xmark & 1 per device & \xmark & Office, Music, Lecture, TV  & Desk, Pocket & \color{black} \mbox{$EER=0.14$}                         \\ \hline
\color{black} \cite{Truong2019_percom}& \color{black} Audible Sound & \color{black} Room Impulse Resp. & \color{black} \checkmark & \color{black} \xmark & \color{black} 1 per device & \color{black} \xmark & \color{black} Office, Kitchen, Corridor, Classroom, Rooms, et al. & \color{black} Room area & \color{black} $ FP,TN \in \{0.089-0.106\}$\\
\hline
\color{black} \cite{Choi2018} & \color{black} Audible Sound & \color{black} Correlation (Sim. Score) & \color{black} \xmark & \color{black} \xmark & \color{black} 1 per device & \color{black} \xmark & \color{black} Office, Parking & \color{black} Personal, Benign & \color{black} \mbox{$EER=0.001$} \\ \hline
\cite{Han2018}       & Ultrasound & Euclidean Distance & \checkmark & \xmark & $m$ per device   & \checkmark & Office & N/A & \color{black} \mbox{$ER=0.02$} (at 50cm)                       \\ \hline
\end{tabular}
\\
\caption{Comparison between \ac{TFA} approaches based on short-range audio channels.}
\label{tab:tfa}
\end{table*}
\end{center}

\textcolor{black}{
\subsection{Lessons Learned}
\label{sec:tfa_lesson}
Our study, summarized by the comparison in Tab. \ref{tab:tfa}, highlights some important \emph{lessons learned}, discussed in the following.
\begin{itemize}
\item \textbf{Reliance on Additional Communication Technologies.} First, we notice that a common feature of all the discussed approaches is to couple the audio channel with another (trusted) communication technology: this is used to transfer the audio recording to the device that is in charge of the comparison, and this step cannot be avoided or replaced. In some cases, such as in \cite{Truong2019_percom}, the auxiliary communication technology is used also to protect against classical attacks against authentication protocols, such as \ac{MITM} attacks.
\item \textbf{Protection Against Relay Attacks.} The schemes discussed in Section \ref{subsec:tfa_literature} are based only on the use of the audio channel. Thus, the protection against MITM attacks, such as relay attacks, is typically delegated to the additional communication technology used to exchange data between the two devices. Alternatively, the actual literature also provides several \ac{MFA} schemes, combining the audio channel with additional sensing modes. As described by the authors in \cite{Truong2014}, pure audio-based schemes could be vulnerable against \emph{relay attacks}, where an active adversary relays messages between two users, cheating them about the sharing of the same physical context. In this scenario, combining the audio channel with further sensing modes, such as WiFi, GPS, and Bluetooth could increase the co-presence assurance, and thus, the robustness of authentication schemes. An example in this direction is the work by the authors in \cite{Truong2014}, where the authors experimentally demonstrated that combining multiple sources improves the robustness of authentication schemes if compared with the use of the audio-only mode. By using the \ac{MCC} statistical measure (with values between $-1$ and $1$), the audio-only mode achieved values of $0.715$, while fusing it with Bluetooth and GPS improves the performance up to values of MCC of $0.978$. In the following contribution in \cite{Truong2015}, the same authors integrated their solution in a real-world application, namely \emph{BlueProximity}, and evaluated the effectiveness of the resulting system against relay attacks.
\item \textbf{Privacy Leakage.} We also remark that, while being effective and valuable approaches, \emph{Listening-Watch}, \emph{Sound-Auth}, and \emph{Proximity-Proof} trade-off protection against active attacks with privacy. Indeed, moving the generation of the sound on the server enables it not only to assess the similarity of the audio recordings, but also to further use the collected audio recordings. Potential privacy leakages can emerge if the server is curious, i.e., it looks for sounds leading to the identification of the ambient where the user is located, extracting sensitive information. We also highlight that potential privacy leakages are increased when multiple sensing modalities are used.
\item \textbf{Performance Issues.} Finally, as experimentally demonstrated by the authors in \cite{Fomichev2019}, the effectiveness of audio-based TFA schemes is strongly dependent on the particular physical context where they are applied. For instance, considering the \emph{Sound-Proof} scheme proposed in \cite{Karapanos2015}, the authors in \cite{Fomichev2019} discovered very different values of \acp{EER} not only in different physical contexts (e.g., car, static office, and office with mobile heterogeneous devices), but also when each physical context is in a different condition. For instance, when deployed in a car, \emph{Sound-Proof} achieves an equal error rate of $0.071$ in a city scenario, while its performance drops down to EER values of $0.124$ when the car is parked, by using the same setup of the protocol parameters. Thus, the deployment of audio-based TFA schemes should be carefully evaluated by the system administrator based on the particular operational context, and its main parameters should be carefully tuned to maximize its performance. 
\end{itemize}
}

\section{Audio Channels for Secure Device Pairing}
\label{sec:pairing}

In this section, we shed lights on the use of audio channels for secure device pairing between devices. The background on the device pairing technique is provided in Section \ref{subsec:back_pair}, while the literature overview and the comparison between the discussed approaches are provided in Section \ref{subsec:pairing_lit}.

\subsection{Background}
\label{subsec:back_pair}

The research area widely known as \emph{Secure Device Pairing} refers to the authentication of unfamiliar devices, i.e., the bootstrapping of secure communication between two wireless and possibly constrained devices, in a way to be robust against eavesdropping and \ac{MITM} attacks~\cite{Halevi2010}. Common pairing operations are, e.g., the association of a Bluetooth mouse, keyboard or headset with a laptop or another communication equipment, or the coupling of a smartwatch with a handheld device. 

The best-known approach to solve such a problem is the standard \ac{DH} protocol. In the DH scheme, two entities sharing no prior secrets can establish a shared key resorting to exponentiation operations~\cite{Diffie2006}. However, because of the lack of any shared knowledge, the authentication of involved parties is missing, and thus, \ac{MITM} attacks are possible. This is the reason why several enhancements of the \ac{DH} protocol have been proposed, leading to authenticated \ac{DH} schemes~\cite{Mirzadeh2014}.

However, in the context of \emph{Secure Device Pairing}, traditional cryptography-based solutions are typically unsuitable. Indeed, the involved devices do not share anything: there is not any shared secret, no \ac{CA} nor \ac{TTP} to leverage for assisted operations. As suggested by the authors in~\cite{Soriente2008}, a theoretical solution perfectly suitable for this case would be the establishment of a common standard \ac{PKI}, trusted by all the devices. However, this would require all manufacturers to agree on a set of features, as well as endless revision processes, leading to large manufacturing delays. Thus, the human involvement in the process is necessary to confirm that the two devices are indeed communicating, and no other party is impersonating one of the legitimate entities. Given the cited premises, 
a challenging problem in this context refers to minimize the effort of humans in pairing operations. 

Many solutions are available in the literature, already summarized by the authors in~\cite{Fomichev2018} and~\cite{Mirzadeh2014}. 
Common solutions are based on \ac{OOB} channels, such as a camera or infrared links. However, these schemes could face deployment issues in some specific contexts, e.g., when the users are visually impaired, there are adversarial ambient light conditions, the area is security-sensitive (military areas) or, finally, the involved devices are not equipped with the necessary external modules, such as a camera or an infrared module.

In these scenarios, the usage of the audio channel could provide enhanced and peculiar features, emerging as a complimentary OOB channel to achieve a secure association between unfamiliar devices.

\subsection{Literature Review}
\label{subsec:pairing_lit}

The idea of using the audio channel for secure device pairing was launched by the authors in~\cite{Goodrich2006}, through the \emph{Loud and Clear (L\&C)} scheme. It uses audio to provide human-assisted device authentication, relying on the use of the spoken natural language. Specifically, \emph{L\&C} covers four possible use-cases, three of them are based on the emission of an audible and syntactically correct sequence by at least one of the devices. If the other entity involved in the pairing process has a speaker, too, it reproduces the detected word and the human has only to compare the two sequences, and to decide if they represent the same word. In case the other device does not have a speaker, the corresponding word is displayed as a text on a screen, and the user has to compare the text with the emitted sound. To reduce the number of audio channel accesses, \emph{L\&C} represents the authentication object as a syntactically correct text (usually non-sense), hashed to generate a fixed bit-string of $H$ bits. Then, the hash is divided into 10-bit sections, and each section is mapped on a word. Thus, $H/10$ words are emitted, resulting in a completion time of about 32 seconds. Even if the scheme leverages an alternative communication channel between the two devices, no security assumption is made on this channel. Note also that the audio communication channel is not secret nor authenticated: thus, eavesdropping is always possible. The assumption about the underlying secrecy of the audio communication channel is instead at the basis of the proposal in~\cite{Halperin2008}, using a low-frequency audio channel to perform the pairing between an \ac{IMD} and a tag reader sharing an unsecured \ac{RFID} connection. Given the audio channel is assumed to be a secret as well as authenticated OOB (AS-OOB), the key is transferred directly via audio in approximately 2 seconds, leading to acoustic eavesdropping attacks if a microphone is placed very near to the human body.

\textcolor{black}{
Starting from the proposal by \cite{Goodrich2006}, a large number of audio-based approaches were proposed for devices pairing.}

Another pairing protocol for constrained wireless devices was proposed in~\cite{Prasad2008}, with reference to devices featuring at least a screen and a microphone. The proposed approach, namely \emph{Beep-Blink}, is based on the comparison of audio-visual patterns at the two sides of the communication to achieve secure device pairing. One of the two devices, featuring audio generation capabilities, encodes via audio a first Short Authenticated String (SAS) and an audio termination string, received by the second device. The other device acknowledges the reception of the two sounds through the blinking of a green led for the first string and a red led for the termination string. The user has just to compare the approximate matching of the two actions, i.e., beeping and blinking, to establish the association between the two devices. Indeed, the security of the scheme is based on the underlying security of the SAS strings.

A simple yet effective technique for secure device pairing on smartphones is provided in~\cite{Peng2007} and further refined in~\cite{Peng2009}. These contributions proposed the \emph{Point\& Connect} scheme, namely \emph{P\&C}, suitable for the pairing of smartphones. Specifically, it leverages a collaborative scheme based on a distance measure between involved devices based on acoustic signals, thus requiring only a speaker and a microphone. Indeed, in a scenario where multiple devices coexist, \emph{P\&C} allows for the selection of the target device through an intention-based mechanism where the initiator shakes and points his/her device towards the selected target destination device. The pairing intention is captured via an acoustic-based distance measuring technique, where the initiator emits two sequential chirp sound signals. At the receiver side, the intended receiver will be the device exhibiting the largest distance reduction, as it is on the \ac{LoS} to the initiator's position. This can be detected via the \ac{RSSI}, pairwise ranging techniques or solutions based on the \ac{ToA}. The proposed technique exhibits outstanding performance (error rate 0) when the positioning of the audio source is at an angle $\alpha=0$ to the receiving device. Otherwise, the performance degrade quickly.

The audio channel is also employed in~\cite{Claycomb2009}, as an OOB channel to assist pairing using an unsecured wireless communication channel. In the proposed approach, namely \emph{UbiComp}, the short-range audio channel is used to convey information to verify cryptography elements delivered on the other, unsecured, wireless channel, essentially to guarantee the physical proximity of the end-device emitting the audio signal. A total number of 2 messages are delivered by the target device on the audio communication channel, consisting of a random number and a public key, and requiring 1.5s in total to be completely delivered, assuming a RSA key pair of 1024 bit. Of course, MITM attacks are possible, given that the public key of the devices is not authenticated. This is why the human is involved in the loop to verify that the sound is effectively emitted by the intended remote device.

Another solution leveraging the audio channel for pairing is the \emph{HAPADEP} protocol~\cite{Soriente2008}. It consists of two consecutive phases. In the transfer phase, triggered when the user pushes a button, one of the two devices transmits its cryptography materials (e.g, the public key) to the other one. Receiving this information, the remote device transforms the key in an audio sequence and plays it. The other device records and decodes the audio sequence, accessing the key. The audio transmission lasts for 3 seconds and, if the pairing is bidirectional, the scheme is repeated with inverted roles. During the following Verification Phase, the user is called to verify that the two sounds emitted by the involved devices are the same, and to press a key if a match is verified. 

Extending the approach in~\cite{Goodrich2006}, the same authors in~\cite{Goodrich2009} proposed a variant based on pure audio, avoiding the need for an auxiliary wireless channel. The basic idea leveraged by this technique is that if an attacker enforces a \ac{MITM} attack by encoding a self-generated string, the legitimate user can detect the audio source and stop the execution of the protocol. 

However, the contributions~\cite{Hu2017} and~\cite{Hu_TDSC2018} confuted experimentally the idea that \ac{MITM} attacks on the audio channel can be always detected because of the human perception of the communication channel.
Hence, in~\cite{Hu2017}, the authors demonstrated that systems based on self-jamming cannot fully reliably detect the presence of signals injected by a malicious adversary that is co-located with the victim, i.e., they are not able to face the so-called overshadowing attack. It is worth noting that facing the overshadowing attacks is more challenging than dealing with co-located MITM attacks described in Section \ref{sec:tfa} in the context of TFA techniques. Indeed, while solutions in the context of TFA can leverage shared secrets or certified fingerprints, this is not possible for the pairing scenario, where the devices are unknown to each other. 

Further proofs of the effectiveness of the overshadowing attack are discussed in~\cite{Hu_TDSC2018}, where the authors also propose some countermeasures. The most effective is based on some security symbols, embedded within the transmitted data at a random location, such that the bit-string size of a generic security symbol corresponds to a single symbol in the bit-stream. For each symbol period, the samples above a given threshold represent security symbols, and they will be verified and discarded by the receiver. The remaining ones will be considered as data, to be demodulated and decoded. Thus, if an attacker transmits further data, it is very likely that the signal energy over the particular symbol will overcome the threshold and it will be discarded, causing a meaningless Denial of Service.

It is worth noting that, contrary to \ac{TFA} schemes discussed in Section \ref{sec:tfa}, in the case of the audio-assisted pairing approaches discussed above the type of audio signal involved in the process is in the audible audio range. This is needed to allow the external user to confirm the correctness of the process. Contrariwise, some other approaches got rid of human involvement. As for the case of~\cite{Peng2009}, a different strategy was proposed by the authors in~\cite{Sigg2012}, that aimed at eliminating the human from the loop. It proposed \emph{AdHocPairing}, an audio-based spontaneous device pairing application leveraging ambient sounds for secure key generation. The two devices synchronously record the ambient sounds for a duration of 6375 milliseconds, and they extract a fingerprint of such a record by taking its \ac{FFT} in 33 sub-bands. To align the registration of the ambient sounds in two close but different locations, they map the recorded audio fingerprint on the code space of an error correction code, using the widest known principle of the Hamming Distance to identify the code-word that best matches the recorded audio fingerprint. One of the two devices, elected as the transmitter, use the best-matching key to encrypt a message, and sends it to the receiver. The receiver, having calculated the key employing the same principle, is then able to decrypt the information. In case it fails, it continues for up 10 attempts, taking every time the best key from the remaining ones having the shortest Hamming distance. While eliminating the human from the loop and resorting to the ambient audio-only, this scheme is more exposed against \ac{MITM} attack, trading off security for usability. In addition, it is defined only for bi-directional pairing, while its specification in the context of unidirectional communications is not defined.

\textcolor{black}{
The authors in \cite{Schurmann2013} provided a thorough study on how the audio channel state recorded by the devices can be used to generate a shared cryptographic secret between two devices in proximity, with minimal guessing probability by third-parties. Specifically, they proposed to extract audio fingerprints from synchronized recordings of mobile devices, to correct their hamming distance from specific pre-defined code-words using fuzzy cryptography schemes, and then to utilize these fingerprints as the seed for cryptographic keys generation among the involved devices. To provide a sufficient level of entropy to the key to be generated, this approach splits the full audio recordings, lasting 6.375 s, into 17 frames of identical length. Then, the \ac{FFT} of each frame is computed and further divided into 33 frequency bands, where the respective energy values are computed. Overall, this operation results in a 512-bit fingerprint. Despite the effectiveness of the described approach, a 100 \% match among fingerprints generated by different devices is very hard to happen. Indeed, the involved devices are characterized by different hardware, are separated by a small distance and they are not perfectly synchronized, thus introducing discrepancies in the fingerprint generation process. Thus, a fuzzy commitment scheme is implemented via Reed-Solomon codes, to generate a common secret among the involved devices.  Using the combination of these strategies, the authors demonstrated how the ambient fingerprinting can be used for fingerprint-based authentication, even in the presence of an active attacker, able to disrupt the shared audio channel inserting audio messages on purpose. 
Specifically, the authors found that a minimum amount of $75\%$ of correct bits in the reconstructed fingerprint are necessary to pair two communicating devices while, at the same time, providing sufficient security guarantees against eavesdroppers experiencing different audio contexts.
A vast experimental campaign is provided to confirm the theoretical results, including office scenarios, situations in which attackers easily imitate the audio environments, and other busy environments.}

\textcolor{black}{
Small variations of the approach in \cite{Schurmann2013} have been described by the authors in~\cite{Nguyen2012} and \cite{Nguyen2018}. The contribution in~\cite{Nguyen2012} focused on the practical issues related to hardware, environment, and time synchronization, necessary to achieve secret key generation via ambient audio fingerprints, and it provided several alternative options for features generation, as well as experimental results. The authors in~\cite{Nguyen2018}, instead, exploited speech recognition techniques on free-form spoken interactions between the owner and the target device to identify precisely the devices to pair. Further, they restrict the pairing only to devices in proximity. The protocol was specified in two forms, based on the presence (or not) of a central authority, and it took up to a maximum of 4 seconds to complete, depending on the speed and length of the sentence used by the owner.}

\textcolor{black}{
While previous approaches only rely on audio, another class of approaches extend the same concept over the whole \emph{context} shared by co-present devices, coupling the audio channel with additional sensors readings.}

\textcolor{black}{
The authors in \cite{Miettinen_ccs2014} introduced the concept of \emph{audio fingerprinting}, i.e., the unique ambient features characterizing a particular physical environment over time. Specifically, the approach proposed in this contribution combined audio-based key generation and a sensor-based approach, using the luminosity features of the surrounding environment. This concept is strengthened by the relationship with the time dimension: in fact, two devices can complete the pairing process only if the \emph{ambient fingerprint} computed over the audio and the luminosity context is similar over a long time-span, i.e., multiple consecutive acquisitions. This innovative concept, namely \emph{Sustained Co-Presence}, is integrated with a fuzzy commitment scheme, that can lead the two devices to estimate the same shared key only if, after the first initial pairing, they continue to share the same ambient context over the time. We notice that the scheme leverages an auxiliary secure communication channel to successfully and securely perform the fuzzy commitment scheme. In addition, acoustic eavesdropping is possible and effective to partially break the protocol, even if it should be coupled with the eavesdropping of the \emph{luminosity} level in the underlying physical environment. The authors of this contribution report similarity values up to $91.8$\% for the audio channel, and of the $95.0$\% for the luminosity sensor readings, making the proposed solution an effective tool for the pairing of wearable and IoT devices.}

\textcolor{black}{
The combination of ambient sound and light have been used also in \cite{DongLiu2017}, where the authors proposed a pairing scheme for wearable devices not equipped with external interfaces. Similarly to the proposal in \cite{Miettinen_ccs2014}, the scheme is rooted in the comparison between ambient sound and light features of the communicating devices, and it does not require strict time synchronization. As a distinguishing feature, the scheme proposed by \cite{DongLiu2017} is triggered only when the user explicitly approves it, limiting the effectiveness of context guessing attacks.  The protocol proposed in this contribution could achieve \acp{EER} of 0.1276 by using only ambient sounds, being however very susceptible to the underlying context features.}

\textcolor{black}{
We recall that the actual literature provides a variety of context-based pairing protocols, possibly implicitly using also the short-range audio channel as an element of the \emph{context} experienced by pairing devices. To provide a reference example, the ambient sound is used by the authors in \cite{Han2018_sp} as part of the features characterizing the physical environment where pairing is carried out. The protocol proposed in this contribution, namely \emph{Perceptio}, does not specifically leverage the audio channel, as it allows two devices that do not share any common sensor to establish a shared key leveraging the physical context, removing any synchronization delay. Given that this survey is focused on short-range audio channels, and the research area of \emph{context-based pairing} is just tangent to this contribution, we refer the interested reader to the surveys in \cite{Sigg2011} and \cite{conti2018survey} for an exhaustive list and comparison of \emph{context-based} pairing approaches.}

The techniques discussed above are summarized in Tab. \ref{tab:pairing}, where differences in key characteristics of the approaches are highlighted

\begin{center}
    \small\addtolength{\tabcolsep}{5pt}
\begin{table*}[htbp]
\begin{tabular}{|P{0.6cm}|P{1.cm}|P{1.6cm}|P{1.0cm}|P{1.1cm}|P{1.5cm}|P{1.89cm}|P{1.2cm}|P{1.1cm}|}
\hline
\color{black} \textbf{Scheme} & \textbf{Feature} &\textbf{Suitable for Unidirect. Pairing} & \textbf{Need of AS-OOB} & \textbf{Audio Channel Usage Time [s]} & \textbf{Acoustic Eavesdropp. Feasibility} & \textbf{Subject to Overshadowing} & \textbf{Auxiliary Secure Channels needed} & \color{black} Perf.\\
\hline
\cite{Soriente2008} & Audible Sound & \checkmark  & \xmark & 3 & \checkmark  & \checkmark & \xmark & \color{black}  \mbox{$ER< 0.05$}\\
\hline
\cite{Goodrich2006} & Audible Sound & \checkmark & \xmark & 32 & \checkmark  & \checkmark & \xmark & \color{black} \mbox{$ER=0$} (within 10~cm)\\
\hline
\cite{Halperin2008} & Audible Sound & \checkmark & \checkmark & 2 & \checkmark  & \checkmark & \xmark  & \color{black} N/A\\
\hline
\cite{Prasad2008} & Audible Sound & \xmark & \xmark & 1 & \checkmark & \checkmark & \xmark & \color{black}  \mbox{$ER< 0.04$}\\
\hline
\cite{Peng2009} & Ultrasound & \checkmark & \xmark & 1.4 & \checkmark & \checkmark & \xmark & \color{black}  \mbox{$ER=0$}  \mbox{(at $\alpha=0$)}\\
\hline
\cite{Claycomb2009} & Audible Sound & \checkmark & \checkmark & 1.5 & \checkmark & \checkmark & \xmark & \color{black}  \mbox{$ER=0$} (within 30~cm)\\
\hline
\cite{Goodrich2009} & Audible Sound & \checkmark & \xmark & 3.4 & \checkmark & \checkmark & \checkmark & \color{black}  \mbox{$ER=0$} (within 30~cm)\\
\hline
\color{black} \cite{Sigg2012}, \cite{Schurmann2013}, \cite{Nguyen2012}, \cite{Nguyen2018} & \color{black} Ambient Sound & \color{black} \xmark & \color{black} \xmark & \color{black} 6.575 & \color{black} \checkmark & \color{black} \checkmark & \color{black} \checkmark & \color{black} $ER <0.001$  \mbox{(indoor env.)}\\
\hline
\color{black} \cite{Miettinen_ccs2014} & \color{black} Ambient Sound & \color{black} \checkmark & \color{black} \xmark & \color{black} $f \cdot w$ per device & \color{black} \checkmark & \color{black} \checkmark  & \color{black}   \checkmark   & \color{black} $EER=0.12$  \mbox{(office env.)} \\ \hline
\color{black} \cite{DongLiu2017} & \color{black} Ambient Sound & \color{black} \checkmark & \color{black} \xmark & \color{black} 6  & \color{black} \checkmark & \color{black} \checkmark & \color{black} \checkmark & \color{black} $EER=0.1276$\\ \hline
\end{tabular}
\\
\caption{Comparison between Pairing Protocols leveraging Audio Channels.}
\label{tab:pairing}
\end{table*}
\end{center}

\textcolor{black}{
\subsection{Secret Sharing via Short-Range Audio Links}
\label{sec:sec_sharing}
When the pairing between two or more devices can leverage a secret, various heterogeneous techniques can be used. We hereby mention a few techniques using the audio channel to either share or allow the computation of a secret.
Specifically, secret sharing is a well-known technique in the literature, used to distribute a secret between some entities in a network, such that subsets of these entities could decode and obtain the secret. For instance, if $n$ is the total number of information pieces (namely, \emph{shares}), and $k$ is the minimum number of shares required to re-assemble the original information, the scheme is said to be a $(k, n)$ threshold secret sharing scheme~\cite{Beimel2011}. 
In this area, the authors in~\cite{Desmedt1998} further introduced \ac{ASSS} techniques, leveraging short-range audible links to achieve secret sharing. These techniques were further studied by the authors in~\cite{Ehdaie2008}, proposing a new, secure but ideal audio secret sharing scheme. The main difference between a regular secret sharing scheme and an audio-based one as the one proposed by~\cite{Ehdaie2008} lies in the numbness of human ears towards phase rotation of the audio signals. This property is used to design a scheme where no computation is required on the receiver side, i.e., on the human ear. The scheme is also demonstrated to be secure, in the sense that exactly $k$ shares are required to reconstruct the signal.
This work was further optimized by the authors in~\cite{Yoshida2012} and~\cite{Washio2014}, concerning the security guarantees and the probability distribution used in the encryption of the original audio secret, respectively. 
In detail, it improves the result of previous works so that shares in ASSS schemes encrypting audio secrets should have bounded amplitude with probability 1. To achieve this objective, the authors use a normal distribution over a bounded domain, and evaluate the security of the resulting scheme. Finally, it is also worth mentioning the recent contribution in~\cite{Miura2016}, further improving the state of the art by estimating the mutual information between secret and shares in a $(n,n)$-threshold \ac{ASSS}.}

\textcolor{black}{
\subsection{Lessons Learned}
\label{sec:pairing_lesson}
The comparison summarized in Table \ref{tab:pairing} highlights several outcomes and trade-offs, each discussed below.
\begin{itemize}
    \item \textbf{AS-OOB Unsuitability.} Legacy approaches leveraging only the audio channel for secure device pairing, e.g., \cite{Halperin2008} and \cite{Claycomb2009}, were based on the AS-OOB assumption. However, the diffusion of audio-based pairing protocols and the large availability of computational resources on-board of modern devices rendered this assumption simply unfeasible. Thus, modern audio-based pairing protocols assume that, when used, the audio channel is accessible also to parties other than the legitimate ones.
    \item \textbf{Acoustic Eavesdropping Vulnerability.} As shown by the summary in Table \ref{tab:pairing}, due to the open nature of the audio channel and the unsuitability of secret sharing, all the described approaches are vulnerable to acoustic eavesdropping attacks. Being passive and stealthy, the adversary could simply listen to the audio channel and, being in proximity of the involved devices, guess a large part of the bits of the secret used for device pairing. This vulnerability, along with the others described below, motivated the deployment of secure auxiliary channels between the pairing devices.
    \item \textbf{Overshadowing and Active Attacks Vulnerability.} As shown by the summary in Table \ref{tab:pairing}, the overshadowing vulnerability discussed in~\cite{Hu_TDSC2018} is an issue for all the approaches, independently from the human involvement in the pairing process. Especially in systems based on self-jamming, the overshadowing attack could be deployed by skilled adversaries, leading to the poisoning of the communication channel and a Denial of Service (DoS) on the pairing process. The authors in \cite{Hu_TDSC2018} also described some possible solutions to thwart overshadowing attacks, including physical-layer solutions based on Frequency Hopping Spread Spectrum (FHSS) and Direct Sequence Spread Spectrum (DSSS) techniques, and MAC-layer ones, based on the inclusion of security symbols within the legitimate audio signal. However, the deployability of these solutions within secure device pairing protocols is not straightforward, as all of them require the sharing of secrets between the involved devices (either for the synchronization of the frequency to be extracted among the possible ones, or for the positioning of security symbols within the audio data stream). The unsuitability of secrets sharing also creates the possibility of other attacks, such as relay and distance hijacking attacks. Relay attacks, being a kind of MITM attacks, assume an adversary that relays messages between the two involved parties, letting them believe to share the same audio context. Similarly, in distance hijacking attacks, the adversary jams the legitimate communication channel between the involved devices after the sharing of the information about the physical context, and completes the pairing process on behalf of one of the legitimate devices. These vulnerabilities, along with the acoustic eavesdropping previously highlighted, motivated the deployment of secure auxiliary channels between the pairing devices. 
    \item \textbf{Secure Auxiliary Channels.} Pairing operations between unknown devices should not leverage any secret pre-shared between involved parties. At the same time, the security of the audio channel could be broken by the passive or the active attacks described above. Therefore, most of the available solutions leverage secure auxiliary communication channels to improve the overall robustness of the pairing scheme. Thus, in the first phase, the devices use the highest layers of the protocol stack and the auxiliary channel (Bluetooth, WiFi, and others) to establish a secure communication link. Then, the co-presence is established using the audio channel. 
    \item \textbf{Hardware Heterogeneity.} In real deployments, the devices taking part in the pairing process could be very heterogeneous, being different not only in the number and the nature of embedded sensors, but also in the hardware chips providing sensing capabilities. These considerations are at the basis of several context-based pairing approaches, relying on physical properties that are invariant compared to the configuration of the pairing device. In this context, the microphones sensing the state of the audio channel could provide a different \emph{intensity} and \emph{timing} of a particular event. Thus, pairing approaches leveraging the audio channel should be able to remove time and intensity differences between pairing devices.
    \item \textbf{Usability.} Most of the analyzed approaches experimentally demonstrated that comparing the audio signals over an increased time improves the performance of the solution, reducing the overall error rate. However, increasing the audio channel usage time leads to an increase in the time needed for the pairing process, reducing the overall usability and transparency of the solution on the users' side. Thus, a trade-off is necessary between security and usability.
    \item \textbf{Additional Sensing Sources.} As also highlighted in the discussion related to TFA schemes, the actual trend is to leverage not only the audio channel for secure device pairing, but also additional sensing sources, such as light, GPS, Bluetooth, to name a few. The combination of all these data is leading to a switch from a pure audio-based secure device pairing to a \emph{context-based} secure device pairing, where the audio is an element of the surrounding context shared by the devices. Many scientific contributions, such as \cite{Miettinen_2015_ccs}, demonstrated the increased robustness of methods based on multiple sensing sources, through context-based proofs of presence that are hard to guess by adversaries.\\
    Despite this evolution, we believe that the study of pure audio-based schemes is still valuable, as the audio context stands as one of the most important elements defining the \emph{physical context} shared by co-present devices at the time of the pairing. Thus, strengthening the co-presence detection and increasing the robustness against well-known attacks on the audio channel (including eavesdropping and overshadowing) are still crucial requirements for any context-based approach leveraging the audio environment.
\end{itemize}}

\textcolor{black}{
\section{Device Authorization and User Verification via Short-Range Audio}}
\label{sec:discrimination}

In this section, we analyze the techniques available in the literature that use short-range audio to provide device authorization and user verification. Section \ref{sec:auth_motivation} provides the background, while Section \ref{sec:authorization_literature} reviews the current literature and provides final comparisons.

\subsection{Background}
\label{sec:auth_motivation}

Authorization 
is a fundamental security service that deals with the problem of identifying if an entity 
has the right to execute a given function on a given device~\cite{stallings2017cryptography}. 
A large variety of solutions exist in the literature, tailored for each system or network to be secured~\cite{lopez2004authentication}. In this section, we focus on authorization mechanisms leveraging short-range audio links to grant or deny access to a system and physical devices.

A first area where audio is used to achieve authorization is in the user registration or authentication to websites, where it is crucial to distinguish a human being from automated software, namely a \emph{bot}. This is a crucial task in many applications, including registration of users to websites, authorized responses from websites and command executions on actuators. This is a very common issue especially for website developers, where correct identification of humans and bots can achieve a valuable defense strategy~\cite{Conti2016_MITM}.
\textcolor{black}{
Throughout this section, we refer to this issue as \emph{user verification}.}

Traditional mechanisms to distinguish between humans and bots were tailored for user registration to websites, and they are based on \acp{CAPTCHA}, widely used nowadays when registering a new account to a website or performing authentication to a known platform. CAPTCHAs identify humans and bots by challenging them with tests that are quite easy for humans, but hard for automatic devices. Examples include visual CAPTCHAs, based on images or videos, text-based or pointing-based ones, where the user has to click with its mouse in a specific location on the screen~\cite{singh2014survey}. 

Another strategy is to recur to short-range audio challenges. Short-range audio is particularly effective for visually impaired individuals or people with motor limitations, as well as when the verifier is a speech recognition system, whose only interfaces with the external world are a microphone and a speaker.

\subsection{Literature Review}
\label{sec:authorization_literature}

In the context of user registration and authentication to websites, many different audio-based CAPTCHAs have been proposed, both based on human speech and ambient sounds, and a comprehensive overview of these strategies can be found in~\cite{Kulkarni2018}. Without loss of generality, they assume a scenario in which the server is the \emph{verifier}. It generates a complex sound, having features that are easy to be recognized by humans but hard to be identified by machines. The remote party, namely the \emph{prover}, has to prove to be a human by providing an interpretation or decoding of the challenge. The overview in~\cite{Kulkarni2018} tackles most of the solutions where the authentication/registration server is the verifier, and the user is the prover. However, many contributions in the last years have demonstrated the weaknesses of most audio-based CAPTCHA solutions~\cite{Sano2013}. Indeed, as reported in~\cite{Bursztein2011} the difference between human and computers audio capabilities is very small, much smaller than the difference between human and computer visual processing capabilities, and this makes it hard to design audio-based CAPTCHAs (further details will be provided in Section \ref{sec:attacks}). The most successful recent techniques are provided by the authors in~\cite{Meutzner2016}, where the authors proposed two different audio CAPTCHA schemes exploiting differences in auditory perception between humans and computers. The first technique explores the auditory perception of humans, especially when multiple simultaneous speakers occur. Indeed, humans defeat computers in isolating several similar acoustic streams from a mixture of signals, despite \ac{BSS} solutions are even more closing the gap. To leverage this gap in the design of an improved audio CAPTCHA, multiple mixtures of speech signals are incorporated into an audio signal, by partially superimposing different words. To still maintain audio decodability features for humans, a variable time delay is introduced between two different words, and all speech pauses are filled with noise speech by multiple talkers. In the second approach, the authors exploit differences in humans' understanding of the natural language. The words in the CAPTCHA are now produced without any noise, and words are selected such as the probability confusion is minimal. To confuse attackers, artificially generated non-sense speech sounds are randomly inserted in the sequence, and the users are challenged to recognize meaningful words from semantically incorrect speech sequences. Note that both of the proposed techniques do not resort to any other communication channel rather than the audio means. 

Also in the context of voice assistants, authorization is a crucial task. Many commercial voice assistants are available nowadays, including Alexa, Siri, Cortana, and Google Now, to name a few. Indeed, they are extremely convenient in scenarios where other kinds of interaction (visual-based or touch-based) require more effort or are unfeasible, such as in Driving Assistance Tools. Thus, ensuring that only authorized sources can trigger voice assistance tools is cumbersome. To provide a further authorization layer, the authors in~\cite{Feng2017} presented \emph{VAuth}, a system guaranteeing that the voice assistant executes commands coming only from the voice of the owner. At the triggering of the voice assistant, the system collect accelerometer data from a wearable component mounted at the user's side, and it correlates them with signals coming from the microphone, via the \ac{MFCC}. Only if there is a match with the profile of the user's voice, the command is issued.

Another application domain where audio channels are essential is in the context of audio-controlled \ac{IoT} devices.  In this dynamic ecosystem, voice interfaces are often provided to simplify the usage of IoT devices in the environment, e.g., by enabling voice-triggered lights turn on, windows closing or opening, doors locking, and many other automated activities. Despite these devices are usually protected by ensuring the physical proximity of the triggering voice commands, often this is not enough. To protect audio assistants from unauthorized usage, the authors in~\cite{Blue2018_asiaccs} propose an authentication mechanism based on (at least) two microphones deployed randomly in the area. The main idea of the proposed system (\emph{2MA}) is to locate the source of any audio source in the scenario, and to authorize the execution only of voice commands whose source is close to the user's mobile device. This is achieved by combining distance bounding, via standard \ac{DOA} mechanisms, and audio similarity, via a \ac{RSH} function. In this case, another communication channel is used between the IoT device and the Mobile Device to exchange information. Thus, an initialization and key establishment phase are necessary. 

Maintaining the focus on audio-controlled IoT devices, the authors in~\cite{Blue2018_wisec} provided a strategy to protect constrained devices from unauthorized actions triggering. Specifically, the authors tackled the issue of the triggering of IoT devices via other devices in the same area emitting sounds via electronic speakers. To distinguish between a human or an electronic speaker, they identify the presence of the sub-bass frequency excitation typical of any modern speaker. Aside, they also design a system filtering the noise of the environment, and computing experimental thresholds to decide if a human or a device is issuing voice commands. Thus, even assuming the attacker can manipulate audio commands, the electronic speaker will be always detected.

An overview of the approaches discussed above is provided in the following Tab. \ref{tab:authorization}.

\begin{center}
    \small\addtolength{\tabcolsep}{8pt}
\begin{table*}[htbp]
    \begin{tabular}{|P{0.75cm}|P{1.6cm}|P{1.6cm}|P{2cm}|P{1.5cm}|P{2cm}|P{1.5cm}|}
    \hline
    \color{black} \textbf{Scheme} & \textbf{Verifier---Generating Sound} & \textbf{Prover---Interpreting Sound} & \textbf{Distinguishing Feature} & \textbf{Auxiliary Secure Channel Needed} & \textbf{Typical Scenario} & \color{black} \textbf{Perf.} \\
    \hline
   ~\cite{Meutzner2016} & Server & Client & Partial overlapping of words & \xmark & User Registration on websites & \color{black} Word Accuracy up to $90.51$\% \\
    \hline
   ~\cite{Meutzner2016} & Server & Client & Natural Language Processing & \xmark & User Registration on websites & \color{black} Word Accuracy up to $98.49$\% \\
    \hline 
   ~\cite{Feng2017} & Server & Client & Accelerometer Data & \checkmark & Voice Assistants & \color{black} Detection Accuracy up to $97$\% \\
    \hline
   ~\cite{Blue2018_asiaccs} & Client & Server & Multiple Receivers & \checkmark & Voice-controlled IoT devices & \color{black} Detection Accuracy up to $97$\%\\
    \hline
   ~\cite{Blue2018_wisec} & Client & Server & Sub-bass over excitation & \xmark & Voice-controlled IoT devices & \color{black} Detection Accuracy up to $99.95$\% \\
    \hline
    \end{tabular}
    \\
    \caption{Comparison between authorization approaches based on short-range audio.}
    \label{tab:authorization}
\end{table*}
\end{center}

\textcolor{black}{
\subsection{Lessons Learned}
\label{sec:authz_lesson}
The following take-away messages can be extracted from our discussion and comparison.
\begin{itemize}
    \item \textbf{Semantic User Identification via Audio.} As highlighted by several studies, when taking the audio as the reference channel for authorization, differences between humans and bots are very small. When the usage of this channel is unavoidable (e.g., in the case of visually impaired users), the mental skills of humans are key to distinguish the nature of the user. Thus, users are asked to understand valid words within streams of characters, or to extract meaningful information from the audible sound. However, these operations have an impact on the overall usability of the solution, increasing the authorization time and overhead.
    \item \textbf{Authorization via Co-Presence Detection.} To protect audio assistants against unauthorized use, legacy techniques based on co-presence detection are used. Despite limiting the effective usage range of these devices, the integration of such physical-layer solutions helps to assess the effective presence of the entity issuing commands, avoiding relay and further MITM attacks.
\end{itemize}
}

\section{Attacks via Audio}
\label{sec:attacks}
In this section, we analyze several attacks performed through the audio channel. They are divided into active attacks, involving the delivering of audio signals on purpose, and passive attacks, performed by simply passively listening on the audio channel.
In Section~\ref{subsec:motivations} we describe the motivations behind these types of attacks, mainly related to exploiting \acp{VCS}. We show that such attacks aim to achieve remote commands execution, as well as to violate user privacy exploiting the acoustic emissions of some devices.
In Section~\ref{subsec-attacks-literature} we provide a thorough analysis of the most representative works that exploited the audio channel for malicious purposes, dividing them into the two aforementioned categories: active attacks (Section \ref{sub:active} ) and passive attacks (Section \ref{sub:passive} ). 

\subsection{Background}\label{subsec:motivations}
The increasing diffusion of \ac{VCS}, such as \ac{VA}, is changing the way people live, offering several automation functions in both the consumer and the business realms. A \ac{VA} is a software, usually installed on computers/smartphones, using different technologies including voice recognition, speech synthesis, and \ac{NLP} to provide different smart services to the users. Thanks to a \ac{VUI}, users can access the VAs and interact with them easily and quickly. 

\begin{figure}[htbp]
    \centering
    \includegraphics[width=\columnwidth]{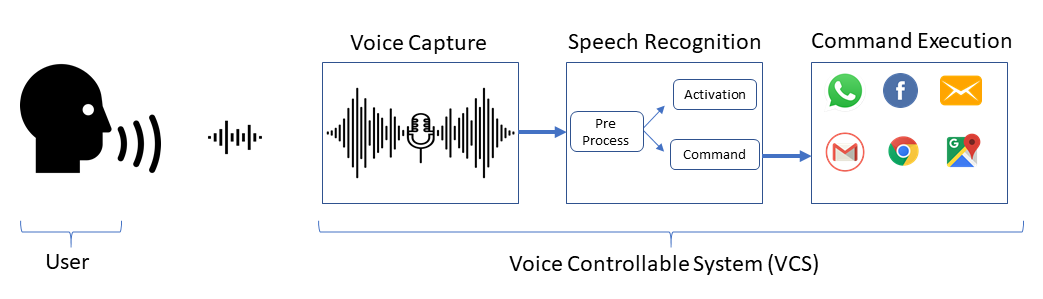}
    \caption{Typical architecture of a VCS.}
    \label{fig:vcs_arch}
\end{figure}

As shown in Fig.~\ref{fig:vcs_arch}, the architecture of a \ac{VA} includes (at least) three main components: voice capture, speech recognition, and command execution. The voice capture module records ambient and speech sounds, that are usually pre-processed and provided to the speech recognition system. This module can identify human voices containing possible commands, isolating them from ambient and noise sounds. Finally, the recognized commands are executed by the command execution module~\cite{Zhang2017}.

A \ac{VA} usually works in two different stages: activation and recognition. The activation stage envisions that the VA 
continuously records ambient sounds, looking for a human voice. Indeed, the user activates the VA by pronouncing a predefined sentence (such us ``Hey Google'', or ``Alexa''). If the predefined sentence is recognized, the VA enters the recognition mode. Other activation methods are possible, e.g., by pressing a physical button, or by opening a specific application. Once entered in recognition mode, the VCS converts voices into commands, thanks to \ac{NLP} techniques.

The VCS technology is constantly evolving, and the \ac{VA} market is progressing at an even higher pace. Indeed, the widespread adoption of \ac{VA} is a key factor behind the enormous growth of smart home devices and applications, with over 275 million \ac{VA} devices expected to control smart homes by 2023, compared to the 25 million estimated in 2018~\cite{juniper_SmartHome}.

Following the rapid increase in popularity, \acp{VA} are gaining more and more attention from industry and academia, raising also important security concerns. Indeed, despite the undoubted usefulness and practicality offered by a voice interface, the lack of any inherent authentication mechanism is a cause of threats to the user's security and privacy, especially when using earlier versions of these softwares.

\subsection{Literature Overview}
\label{subsec-attacks-literature}
The active attacks discussed in the recent scientific literature are mainly focused on attacking the VCSs. Indeed, these active attacks can be performed by sending audio signals on purpose to VCSs. These signals are neither unintelligible nor inaudible to humans, but still recognizable by voice capturing subsystems. Section~\ref{sub:active} provides a description of the most important contributions in this field, while Tab.~\ref{tab:active} summarizes and compares them.
The most representative passive attacks discussed in the literature are described in Section~\ref{sub:passive}. Finally, Tab.~\ref{tab:passive} highlights and cross-compares them across their main features.

\subsubsection{Active Attacks}
\label{sub:active}
The majority of the attacks described below aims to cause the execution of unauthorized commands in a target device equipped with a VCS. These commands include malicious activities, such as data exfiltration (by sending of SMS or emails), money stealing (via payment-based SMS services), malware installation (by opening malicious web pages and starting downloads), and possibly others.

One of the first attacks against a voice assistant has been performed by the authors in~\cite{Diao2014}, which proposed a bypassing attack exploiting the lack of any permission necessary for an Android application to access the phone speaker. The adversary model involves the use of a malware to invoke the voice assistant (Google Voice Search), bringing it to the foreground. Then, the malware plays a pre-prepared audio file containing specific commands, recognized and executed by the VCS. Indeed, many critical assumptions are necessary for the attack to be successfully executed. The victim's smartphone should be already infected with the malware, and the attack could be successfully completed only if the user is away from the phone.  
Indeed, if the victim is close to his smartphone, he can notice the ongoing attack by hearing the audio played by the malware through the speaker. Thus, he could manually block the execution of the command. Moreover, the attack has been tested only in a closed environment, without any external noise.

Another pioneer contribution that explored the feasibility of attacking a VCS using an audio signal unintelligible to humans was discussed in~\cite{191968}. Authors provided a thorough study of the speech recognition mechanisms, investigating if differences in the ways humans and machines detect spoken speech could lead to exploitable vulnerabilities.
They found that these differences can be easily exploited by an adversary to produce sounds that are recognized as a command by VCSs, while being perceived by humans as noise other than speech. This methodology increases the potential danger and the feasibility of these types of attacks. Indeed, these powerful attacks involve neither installing a malware nor performing any further interactions with the attacker and the victim's smartphone before the attack execution. Moreover, the audio signals injected by the attacker to activate and use the VA are stealthy, as they are unintelligible to human listeners. The attacker first produces an audio file containing a regular voice command that she wants to be executed on the targeted device. Then, the audio file is provided as input to specific software, namely an ``audio mangler''. This software removes all the audio signal features not used by the speech recognition module, but used by a human. The authors evaluated the real feasibility of this attack against the Google Now personal assistant, playing the recorded commands using speakers located about 30cm from the targeted device. Their experiments involved several unauthorized actions, including activating the VA (Google Now), calling a number, sending SMS, and opening a website.
Even if this work improved the previous attacks against VCS, some limitations are still present. These limitations involve the audibility of the unauthorized commands and the white box approach, which assumes some knowledge of the victim's speech recognition system by the attacker.

Some of these limitations have been overcome in~\cite{Carlini2016}. In this paper, the authors proposed an attack against any VCSs, using a malicious audio signal coming from outside the victim's smartphone. The attack is realized with a few knowledge about the speech recognition system of the user, and it applies to a variety of scenarios. The attack is performed against the Google Now personal assistant software and the open-source CMU Sphinx speech recognition system, used by several VCSs. The authors significantly improved the state of the art by evaluating the feasibility of this attack under more realistic scenarios. In addition, they formalized a general method to produce machine-understandable speech that is, at the same time, rarely recognized by humans. The attack was performed in an isolated room, without background noise, playing the audio commands on a speaker located exactly 50cm from the microphone. This work practically demonstrated the feasibility to remotely attack a VCS, by activating a virtual assistant software and exploiting it to execute several commands without neither authorization nor privileges in the victim's system. On the opposite side, these attacks use hidden voice commands which are incomprehensible but still audible to humans. Thus, the user might notice the ongoing attack by hearing a noise.

To overcome this limit and increase the reach of the attack, a completely inaudible attack has been proposed by the authors in~\cite{Zhang2017}. This objective is achieved by modulating voice commands on ultrasonic carriers ($f > 20 kHz$), in a way to be inaudible to humans. 
The authors performed an in-depth study on the voice capturing subsystems (composed by microphones), investigating how they record audible sounds. They found that electric components such as amplifiers follow a non-linear behavior compared to the input signal features. Then, they found the same property on real microphone modules, observing that an attacker can exploit it to build an ad-hoc audio signal, containing the command. The voice is then modulated on an ultrasonic carrier right before the transmission. On the reception side, the sound is correctly demodulated on base-band by the receiver hardware. To demonstrate the feasibility of their methodology, the authors validated it using the most important speech recognition systems, such as Apple Siri, Google Now, and others. Specifically, they were successful in performing a variety of actions by simply injecting a sequence of inaudible voice commands, including accessing a malicious website, spying the victim accessing the image and the sound of its device, injecting false information, and many others.

Tab. \ref{tab:active} summarizes our discussion and highlights similarity and differences of the approaches.
\begin{center}
    \small\addtolength{\tabcolsep}{8pt}
\begin{table*}[htbp]
\begin{tabular}{|P{0.5cm}|P{0.8cm}|P{0.9cm}|P{2.0cm}|P{1.5cm}|P{1cm}|P{1.2cm}|P{2.2cm}|}
\hline
                               \color{black}\textbf{Scheme}   & \textbf{Inaudible} & \textbf{Max attack range} & \textbf{Target Software}                                                                                                            & \textbf{Attacker Capabilities}            & \textbf{Attack Source} & \textbf{Tested Scenario}                & \textbf{\textcolor{black}{Performance}}\\ \hline
\multicolumn{1}{|l|}{\cite{Zhang2017}}     & \checkmark       & 165 cm           & Apple Siri, Google Now, Samsung S Voice, Cortana, Amazon Alexa, Huawei Hi Voice & Commands Recognition, Activation & Remote        & Office, Cafe, Street & \textcolor{black}{Attack Accuracy from 30\% to 100\% according to the command}\\ \hline
\multicolumn{1}{|l|}{\cite{Diao2014}}    & \xmark         & Inside             & Google Search App                                                                                                          & Commands Recognition             & Local         & Noiseless Scenario         &\textcolor{black}{Attack Accuracy 100\%}          \\ \hline
\multicolumn{1}{|l|}{\cite{191968}}    & \xmark         & 30 cm             & MFCC-based speech recognition systems                                                                                                          & Commands Recognition, Activation             & Remote         & Quiet room         &   \textcolor{black}{Being a feasibility study, the authors did not provide any performance indicator}        \\ \hline
\multicolumn{1}{|l|}{\cite{Carlini2016}} & \checkmark       & 50 cm            & Google Now, CMU Sphinx                                                                                                     & Commands Recognition, Activation & Remote        & Recorded Background noise samples          &\textcolor{black}{Attack Accuracy over 90\%}    \\ \hline
\end{tabular}
\\
\caption{Comparison between active attacks via short-range audio channels.}
\label{tab:active}
\end{table*}
\end{center}

Among other scientific contributions in this category, it is worth mentioning the work in~\cite{Wu2013}, where the authors provided an overview of the main technologies behind VCSs, including voice conversion and speaker verification. Furthermore, the most important spoofing attacks have been studied in a variety of scenarios, with a focus on the modification of a speaker's voice to emit sounds similar to the ones emitted by a different speaker, without modifying its content (namely, voice conversion spoofing attacks).

\subsubsection{Passive Attacks}
\label{sub:passive}

Acoustic side-channel attacks have been largely used to capture information from different devices, such as keyboards, printers, CPU, and even to identify information written by handwriting. 

These attacks are mainly performed by placing a covert listening device in the physical proximity of the target device. This device simply records acoustic emanations from the victim device. Then, the recorded audio is processed to extract the desired features, further used to train a \ac{ML}-based model. This attack has been proved to be feasible even recording the acoustic emanation through a \ac{VOIP} connection. The main goals include, but are not limited to, the victim's privacy violation, i.e., recovering the typed/printed text, or the violation of the intellectual property, i.e., recovering printed objects.

In the context of acoustic emanations analysis, the authors  in~\cite{Backes2010} presented an attack against dot-matrix printers, intending to recover the printed English text. The attack includes recording the acoustic emissions with a microphone placed at about 10 cm from the targeted device. Indeed, at such a short distance, an attacker can recover up to 72\% of printed words. Better performances, with a recovering percentage of up to 95\%, can be obtained by assuming a contextual knowledge about the printing text. 
The attack can be divided into two steps. It involves a preliminary training phase, where the sound of printed words is recorded and used to train a machine learning-based model, and a second recognition phase, where printed English text is recognized. In the training phase, a word-based approach has been used instead of decoding individual letters, due to decay times and the induced blurring across adjacent letters. Interestingly, the authors found that most of the features characterizing printed sounds are located above the 20 kHz threshold. Thus, they identify the words in the recorded audio, analyzing the power spectral density above the 20 kHz acoustic threshold, and then spread the filter frequencies linearly over the whole bandwidth. Finally, they used digital filter banks to perform sub-band decomposition on each word. The authors tried also to perform the same attack against Ink-jet printers and laser printers, concluding that these printers technologies seem to be unaffected by this kind of attack.
A similar attack against manufacturing systems has been investigated by the authors in~\cite{Faruque2016}. The authors demonstrated that the sound emitted during the creation of an object effectively carries specific information about the process. Indeed, this information can be leveraged to reconstruct the printed item, even without any knowledge of the original design, violating intellectual property. The attack consists of six phases. In the first \textit{Acoustic Data Acquisition} phase, the acoustic emissions of the 3D printer are captured using a microphone placed 20 cm from the target. It is followed by a \textit{Pre-processing} phase, involving the analysis of the audio to remove undesired noise. Then, a \textit{Feature Extraction} phase is executed, where the time and frequency domain features commonly used in speech pattern recognition are selected to training the learning algorithm. Then, a \textit{Regression Model} is applied to determine the speed of the printing. As per the definition, such a model consists of a collection of models each using a supervised learning algorithm for regression. Then, a \textit{Classification Model for Axis Prediction} is used to determine the axis movement, and a \textit{Direction Prediction Model} is applied to detect the direction of motion, using the frame energy of the audio signal. Finally, a \textit{Model Recreation} process is applied to reconstruct the object using the output of the previous phases.
The authors tested and validated their attack against a modern 3D printer, achieving an axis prediction accuracy of 92.54\% and a length prediction error of 6.35\% on a complicated object, such as a door key~\cite{Faruque2016}.

The same type of attack has been largely used also against keyboards. In this case, the attacker aims to recover the typed text, based on the observation that different keystrokes produce different sounds. 
The authors in~\cite{asonov2004keyboard} used neural networks with labeled training samples, identifying keystrokes using the \ac{FFT} method and achieving an accuracy of 80\%. This result has been improved a few years later by the authors in  ~\cite{Zhuang2009}. The authors revisited the attack methodology using unlabeled keystrokes samples and \textit{Cepstrum} features, increasing the accuracy of up to 96\%. A different approach is provided in~\cite{Berger2006}, motivated by the observation that the sound emanated by a keystroke is related to its position on the keyboard. In fact, keys that are positioned close to each other emit more similar sounds if compared to those positioned far away. Moreover, the authors observed that working at the granularity of words instead of single keys, it is possible to leverage the statistical properties of the English language. Following these intuitions, the authors presented a dictionary attack able to recognize single words, having from 7 to 13 letters. Through specific signal processing tools, the authors achieved an accuracy of 73\% over all the tested words. Furthermore, the attack does not require a training phase.

All these attacks are performed placing a microphone in the proximity of the targeted device. Despite this scenario poses a serious threat to the confidentiality of passwords and other sensitive text, the real applicability is limited. Indeed, the attacker should be located in the very near physical proximity of the user, being easily guessable. 

To overcome this limitation, the authors in~\cite{compagno2017} presented a new scenario, considered the acquisition of acoustic information using VoIP protocols. The authors recorded keystroke sounds through a Skype call and demonstrated the feasibility of discovering what a user is typing, using a machine learning classifier based on \ac{MFCC} features. They also investigated the impact that some variables have on the accuracy of their classifier, including the bandwidth and the presence of speech in the recorded audio. 

The same attack is revisited by the authors in~\cite{Anand2018_CODASPY}, with a focus on the recovery of random passwords and PINs. Instead of using a word-based approach as in previous work, the authors presented a methodology focused on single character detection. Thus, the accuracy of the attack depends only on a single character detection rate, avoiding the use of any language model and dictionaries. The training phase consists of the collection of single keystrokes (30 instances per key). Then, the authors used the trained model to detect randomly generated 6-character passwords, containing only lowercase letters (a, z), and 4 digits PINs. This is performed by using MFCC features extracted from password keystrokes, eavesdropped over the remote call. A similar methodology was used by the authors in~\cite{Halevi2015}, where the authors studied the vulnerability derived by audio emanations of keyboard typing. The authors showed that the keyboard eavesdropping attack performance is affected by a few variables, such as the typing style of the particular user, the data inserted by the user, and the adopted detection strategy. 

Another type of acoustic side-channel attack was performed by the authors in~\cite{deSouza2019} against PIN pad devices. This attack, called Differential Audio Analysis (DDA), analyzes the differential characteristics of the sound captured by two microphones placed inside the targeted device. Such a difference is then expressed as the cross-signal transfer function. Their experiments achieved the 100\% of accuracy for certain devices, while only the 63\% of accuracy was achieved when the target device does not produce the sufficient level of audible sound in correspondence to the press of a key.

Finally, a side-channel attack against handwriting has been investigated by the authors in~\cite{Yu2016}. The attack leverages the acoustic emissions of people handwriting recorded through a mobile phone. Indeed, this attack could lead to the leakage of personal information, eavesdropping, e.g, the sound derived from people filling out privacy-related forms. The authors presented a methodology based on audio signal processing and \ac{ML}, able to recover the handwritten text. The audio is recorded with a smartphone placed in the same desk of the target user, about 30 cm far from the writer. Indeed, they achieved an accuracy of about 60\% in word recognition, paving the way for further refinements and improvements.
The approaches described above are summarized and cross-compared in Tab.~\ref{tab:passive}.

\begin{center}
    \small\addtolength{\tabcolsep}{7pt}
\begin{table*}[htbp]
\begin{tabular}{|P{0.05cm}|P{2.cm}|P{1.5cm}|P{1.0cm}|P{1.5cm}|P{1.3cm}|P{1.5cm}|P{1.5cm}|}
\hline
    \color{black}\textbf{Scheme}   & \textbf{Eavesdropp. Position} & \textbf{Attack Vector}   & \textbf{Target}    & \textbf{Main Goal}            & \textbf{Features}                          & \textbf{Algorithm}   & \textcolor{black}{\textbf{Performance}}\\ \hline
\multicolumn{1}{|l|}{\cite{Anand2018_CODASPY}} & Same Room (15 cm)                          & Smartphone's Microphone          & Keyboard           & Password and PIN Detection   & Sequential Minimal Optimization           & MFCC                                                                                           & \textcolor{black}{Attack accuracy 74.33\%}        \\ \hline
\multicolumn{1}{|l|}{\cite{Yu2016}}       & Same Writing Surface (20-30 cm)       & Smartphone's Microphone          & Hand writing        & Word Detection               & SVM Based              & FFT                                                                                                              & \textcolor{black}{Attack accuracy 55\%}    \\ \hline
\multicolumn{1}{|l|}{\cite{Backes2010}}  & Same Room  (10 cm)                    & Sennheiser MKH-8040 microphone & Dot-matrix printer & Recover Printed English Text & Undeclared              & Sub-band decomposition                                                  & \textcolor{black}{Attack accuracy 72\%}        \\ \hline
\multicolumn{1}{|l|}{\cite{Faruque2016}}  & Same Room (20 cm)                    & Condenser microphone Zoom H6   & 3D printers        & Reconstruct Printed Object   & Regression Model           & Frame energy, Zero Crossing Rate, energy entropy, spectral entropy, spectral flux, MFCC & \textcolor{black}{Attack accuracy 92.54\%}         \\ \hline
\multicolumn{1}{|l|}{\cite{deSouza2019}}  & Inside the Target Device                     & 2 microphones   & PIN pads        & PIN detection   & Yule-Walker Auto-regressive Method          &  Signal Transfer Function & \textcolor{black}{Attack accuracy 100\%}           \\ \hline
\end{tabular}
\\
\caption{Comparison between passive attacks via short-range audio channels.}
\label{tab:passive}
\end{table*}
\end{center}

\textcolor{black}{
\subsection{Lessons Learned}
\label{sec:pasvAttacks_lesson}
The discussion on active and passive attacks carried out in the previous subsection allows us to highlight the following lessons learned:
\begin{itemize}
    \item \textbf{Voice activation as a defense Tool.} All attacks against the \acp{VA} described in Section~\ref{subsec-attacks-literature} are based on the delivery of a malicious vocal command to the \ac{VCS} of the target device. Then, the audio signal is interpreted as a legitimate command by the VA. However, to execute it, the attacker first has to activate the VA. The recent versions of common \acp{VA} are equipped with a \emph{voice activation} feature, enabled by default, which allows the attacker to activate the VA with the same methodology, as the microphone is always active. Activating the voice assistant using a physical button, as in the oldest versions, would significantly mitigate the problem, requiring the attackers to use more complex attack mechanisms, e.g., activating the VA through a malware. These attacks are also promoted by the lack of defense techniques protecting VAs, such as using vocal authentication mechanisms, recognizing the voice of authorized users and enabling them only to use the VA.\\
    \item \textbf{Unintelligible/Inaudible audio commands.} As discussed in the previous section, attacks performed through inaudible commands could be very difficult to detect by the legitimate user. In fact, if the malicious signal is properly modulated, the audio command is perceived as noise by humans, while remaining intelligible by speech recognition systems. Moreover, if the attack is performed by using audio signals inaudible by humans, the attack becomes completely stealthy. This attack is enabled by the microphones commonly used on smartphones and other electronic devices, which are sensitive even at frequencies other than those audible to humans. Indeed, many microphones can sense sounds generated at frequencies higher than 20 kHz, even without any use case justifying it.\\
    \item \textbf{Extended attack range.} The first attacks against VA systems were effective only if performed near the target device. This consideration considerably reduces the real use cases of the attacks, making them difficult to execute. Unfortunately, as demonstrated by the authors in~\cite{Zhang2017}, by using the appropriate equipment and technologies, the attack range can be extended to larger distances, including 165cm. This wide distance, together with the inaudible features of the malicious audio command used for the attack, severely increases the severity of the threat. We highlight that, also in this case, the vulnerability is introduced mainly by the microphone, designed with properties higher than those required by common use cases. By decreasing the sensitivity of the microphone, it could be possible to limit the distance between the device and the audio source, without affecting the usability of the VA in regular use cases, usually requiring a distance of a few centimeters between the user and the device.\\
    \item \textbf{Audio as a side-channel.} The attacks described in Section~\ref{sub:passive} demonstrate the feasibility of different kinds of privacy leakage, mainly due to the acoustic signals emitted by target devices. Although in many cases the performances are not very high, and usually some knowledge is needed about the victim's system, the feasibility of these attacks should not be underestimated. The lack of effective countermeasures highlights that further efforts by the research community are needed to mitigate this problem.
\end{itemize}}

\section{Defense mechanisms using short-range audio}
\label{sec:defense}

This section explores the use of audio channels as a defense mechanism against attacks previously described. Section \ref{sec:motivation_defense} provides the background, while Section\ref{sec:lit_over_defense} delves into the current scientific literature.

\subsection{Background}
\label{sec:motivation_defense}

The effectiveness of the attacks discussed in the previous section has motivated the design of ad-hoc defense techniques to limit their impact. 

The idea followed by many authors consists of emitting sounds on purpose to protect the sound carrying information about a specific process. Without loss of generality, two approaches are used. The first is based on white noise emission, where an audio-enabled system is coupled with the target device and emit sounds that are flat in the frequency spectrum and overcome, in volume, the other ones. The second approach, instead, is based on the generation of sounds that are equal to the ones produced by the target device, thus poisoning the information received by the attacker.

The following discussion provides details about the usage of such techniques in specific sensitive conditions.

\subsection{Literature Overview}
\label{sec:lit_over_defense}

A glaring example is provided by the authors in~\cite{Anand2016_wisec} and~\cite{Anand2018_TDSC}, in the context of vibration-based communications. These contributions tackle the PIN-Vibra method~\cite{Saxena2011_percom}, used to transmit the keying material between two devices thanks to an ON-OFF vibration scheme. The receiver, equipped with MEMS motors, can detect the presence of vibrations and decode the key or, in general, a piece of information. While being exciting, these techniques have been demonstrated to be susceptible to eavesdropping attacks exploring the sound generated by vibrations~\cite{Halevi2010},~\cite{Halevi2013_TIFS}. To overcome such attacks,~\cite{Anand2018_TDSC} proposed Vibreaker, a system minimizing the leakage of information from the vibration-based communication channel via white noise masking. Specifically, Vibreaker couples the physical vibration with simultaneous audio emissions via the microphone of the smartphones involved in the communication. White noise is emitted by the microphone to cover sounds emitted by the vibrations, while low-frequency tones are further generated on-purpose and emitted to compensate for the partial inability of smartphone speakers to emit low-bandwidth sounds. \textcolor{black}{ On the attacker side, the best (minimum) Error Rate is 30\%, that is very far to be acceptable for any valuable usage.} Note that, while the user is not explicitly involved in the process (with undoubted usability gains), equipping the devices with a microphone causes a little overhead on the system, requiring not only motor accelerators but also audio-enabled devices. Similar properties were used by the authors in~\cite{Kim2015}, in the context of medical devices. As before, the main communication channel is vibration-based and it involves a medical device and external equipment (smartphone or handheld). The proposed scheme, namely SecureVibe, is secured against acoustic eavesdropping derived from vibrations by generating appropriate masking sounds. As for the previous proposals, despite not requiring any active user involvement, this countermeasure forces a little overhead in the required components, as a speaker should be added (or coupled) with the medical device.\textcolor{black}{ The authors tested the effectiveness of their scheme against the ICA technique, and they verified that the scheme is secure up to a distance of 15~cm between the communicating entities. Instead, when the distance is higher than 15~cm, the ICA technique could be used to isolate the single signals, compromising the effectiveness of the technique.}

Emitting white noise and fake keystrokes sounds are proposed by the authors in ~\cite{Anand2017_FCDS} and~\cite{Anand2018_CODASPY} to protect against attacks on keyboard typing. As thoroughly described in Section \ref{sec:attacks}, the mechanical sound emitted by pressing and releasing keyboard keys can lead to their identification. To provide an effective defense, white noise and fake keystrokes are proposed, and their impact on the capability of the adversary to recover the correct key is evaluated. \textcolor{black}{ The best performance are achieved using fake keystrokes, where the attacker achieves a minimum error rate of 33\%. }

A similar defense mechanism is proposed by the authors in~\cite{Shrestha2017_wisec}, concerning a zero-effort de-authentication system called ZEBRA, proposed by the authors in~\cite{Mare2014}. The system is made up of two components: a smartwatch and a software running on the PC, pre-paired wirelessly to each other. Both of them record the event they observe, including keyboard and mouse interactions, and the software on the PC compares continuously the recorded events, evaluating their matching. When they no longer match, the user is assumed to be far from the PC and the user is de-authenticated. Several contributions, such as~\cite{Hutha2015}, demonstrated that ZEBRA can be cheated by an adversary that mimics the behavior of the user, e.g., via keystroke sounds emitted on-purpose. To provide a defense against such attacks, the authors propose to use the sound masking strategy, obtained by having the login terminals or a device placed in the surrounding environment to produce deliberate sounds, obfuscating the sound coming from the keyboards, and thus limiting the attacker possibility to reproduce the victim user’s activities. \textcolor{black}{ Compared to the legacy ZEBRA scheme, the described defense mechanisms can improve the robustness to the aforementioned attacks up to the 70\% error rate on the attacker side, complicating the effort of the adversary. }

Defense-mechanisms using white-spectrum noise to thwart passive eavesdroppers are also used by~\cite{Zhang2014} and~\cite{Zhang2018}. 
The system proposed in these contributions consists of a sound-based communication channel, where speakers and microphones are used to deliver and receive information encoded in short-range audio messages.
Given that the same speakers and microphones are used to emit the white-spectrum noise, this is the only technique where the defense mechanism can really be defined as \emph{device-free}, as it does not require any additional device if compared to the unsecured scenario. \textcolor{black}{ While in the regular setting of the algorithm the ICA algorithm can isolate the waveforms in 90\% of the cases at a minimum distance of 5cm, using random movements with a speed of 30cm/s the effectiveness of the ICA technique decreases to the 40\%, complicating the task of identifying the correct waveform emitted by the legitimate devices.}

A similar approach was used by the authors in~\cite{Nandakumar2013}. Specifically, the authors proposed a system called Dhwani, i.e., an acoustics-based \ac{NFC} system that uses the microphone and speakers on mobile phones within the same scope of legacy NFC communications, thus eliminating any specialized NFC hardware. As for~\cite{Zhang2014} and~\cite{Zhang2018}, Dhwani uses the \emph{JamSecure} technique, i.e., a self jamming strategy combined with self-interference cancellation at the receiver. This technique allows achieving secure communications from the information perspective among the communicating entities.

It is worth noting that, while the sum of audio signals could be a simple solution to avoid the leak of information towards trivial adversaries, smart and powerful adversaries can theoretically recur to algorithms such as \ac{ICA} to decouple the summed signals, leading to the identification of the single components~\cite{hyvarinen2000independent}. However, some experimental studies performed by the authors in~\cite{Kim2015},~\cite{Zhang2014} and~\cite{Zhang2017} experimentally demonstrated that if the two sound sources are very close to each other---e.g., within few centimeters---, the channel difference cannot be recognized by the receiving microphones, thus nullifying the efficacy of the ICA solution.

Another property of the audio signal that can be leveraged to provide enhanced defense tools is the proximity of two devices. Indeed, a well-known property of the audio signal is its quick attenuation over distance and physical barriers. Thus, the amplitude difference between the two recordings can be used to provide a rough estimation of the proximity of two devices. This principle is used in~\cite{Shretsha_2018_cns} to provide a solution for the Home Alone Wearable (HAW) attack. This attack is based on the assumption that users do not lock their wearable after their use and leave them unattended, hence enabling a malicious adversary to use the wearable to access information on the user device---e.g., a smartphone. To detect physical proximity between the user's smartphone and the wearable, the wearable triggers the phone to produce a sound, and both the devices record it via the microphone. The wearable then sends the recording to the smartphone via a Bluetooth connection, and the particular command issued on the wearable is executed only if the similarity score between the two recording exceeds a given threshold. It is worth noting that the user is not involved in the process, i.e., he/she does not have to state its explicit consent to authorize the information exchange. However, the wearable should be equipped with a microphone and the generated sound should be in the audible range, given that the microphones on the wearables usually cannot sample a tone in the ultrasound frequency range. 

With the same objective in mind and using the same basic concepts, i.e., physical proximity, the iLock technique proposed by~\cite{Li2016} recognizes the user's presence through the evaluation of the physical separation from the personal device, leveraging the changes in the wireless acoustic signals, and it immediately locks the hosting device if a threshold distance is overcome. iLock is based on the calculation of the \ac{ToF} of the high-frequency acoustic signals (ultrasounds) emitted by the speakers. Being designed for smartphones, iLock does not require additional devices, given that microphones and speakers are already available in every commercial device, and it does not involve any further human effort. However, when applied in other devices, such as wearables, it could lead to an increase in the minimum features to be included in the system design. 
\textcolor{black}{ Considering the strongest attacker (i.e., an attacker that is closer to the phone than the legitimate user), iLock achieves a detection accuracy of the 60\% when there is a difference of 0-270 degrees between the orientation of the user and the orientation of the attackers to the phone.}

Physical proximity between communicating devices is also used by the authors in~\cite{Halevi2012_esorics} to defend against Reader-and-Ghost attacks affecting \ac{NFC} devices. In these attacks, a malicious reader tries to involve a legitimate NFC device in an unwanted transaction, piloted by a ghost adversary in another physical location, while the owner of the device wanted to perform a different transaction. To solve this issue, a transaction verification process is required, where the owner of the device specifically approves that transaction, and not a different one. To provide a defense mechanism against such a threat, the authors in~\cite{Halevi2012_esorics} proposed a transaction verification strategy able to assure the proximity between a server and a mobile phone, leveraging the correlation between the related ambient audio recordings. The similarity between the two recordings is established via the computation of the cross-correlation, as described in Section \ref{subsec:audio_comparison}, and it does not require any user involvement. Without loss of generality, the work by the authors in~\cite{Halevi2012_esorics} aims at detecting the co-presence between the involved devices. \textcolor{black}{As discussed in previous sections \ref{sec:tfa}, \ref{sec:pairing}, and \ref{sec:discrimination}, co-presence detection strategies, including context-based strategies based on multiple sensing sources, are used for a variety of security-related tasks, including authentication, pairing, and access control. We refer the interested reader to the recent survey in \cite{conti2018survey} for a comprehensive overview of the strategies using sensing modalities other than the audio channel.}

Tab. \ref{tab:defense} summarizes the above contributions and highlights similarities and differences across the application scenarios.
\begin{center}
    \small\addtolength{\tabcolsep}{7pt}
    
\begin{table*}[htbp]
\begin{tabular}{|P{0.6cm}|P{1.3cm}|P{1.3cm}|P{1.3cm}|P{1.0cm}|P{1.6cm}|P{1.8cm}| P{1.5cm}|}
\hline
\color{black}\textbf{Scheme} & \textbf{Audio Signal} & \textbf{Main Channel} & \textbf{Physical Property} & \textbf{Device free} & \textbf{Explicit User Involvement} & \textbf{Scope} & \color{black} \textbf{Perf.} \\
\hline
\cite{Anand2016_wisec}, ~\cite{Anand2018_TDSC} & White Noise & Vibration & Masking Sounds & \xmark & \xmark & Eavesdropping Protection & \color{black} Attacker Best Error Rate 30\%  \\
\hline
\cite{Kim2015} & White Noise & Vibration & Masking Sounds & \xmark & \checkmark &  Eavesdropping Protection & \color{black} 50\% Error Rate up to 10~cm  \\
\hline
\cite{Anand2017_FCDS}, ~\cite{Anand2018_CODASPY} & White Noise, Fake Keystrokes & Acoustic (Keystroke) & New sounds generation & \xmark & \xmark & Eavesdropping Protection & \color{black} Keystroke Detection Error Rate 33\% \\
\hline
\cite{Shrestha2017_wisec} & White Noise, Music Sounds & Acoustic (Keystroke) & Masking Sounds & \xmark & \xmark & Eavesdropping Protection & \color{black} Attacker Error Rate 70\%\\
\hline
\cite{Zhang2014}, \cite{Zhang2018}, \cite{Nandakumar2013} & White Noise & Acoustic & Masking Sound (Eraseable) & \checkmark & \xmark & Eavesdropping Protection & \color{black} Eavesdropping Success Rate 40\% \\
\hline
\cite{Shretsha_2018_cns} & Fake Audible and Ambient Sound & RF & Proximity Detection & \xmark & \xmark & Anti-Theft & \color{black} Maximum Attacker False Acceptance Rate 5.3\%\\
\hline
\cite{Li2016} & Ultrasound & Physical Contact & Distance Bounding & \xmark & \xmark & Anti-Theft & \color{black} Attack Detection accuracy 60\% \\
\hline
\cite{Halevi2012_esorics} & Ambient Sound & NFC & Proximity Detection & \xmark & \xmark & Transaction Verification & \color{black} Attack Detection Accuracy 100\% \\
\hline
\end{tabular}
\\
\caption{Comparison between defense approaches leveraging short-range audio channels.}
\label{tab:defense}
\end{table*}
\end{center}

\textcolor{black}{
\subsection{Lessons Learned}
\label{sec:defense_lesson}
The most important considerations and take-home messages arising from the above discussion are summarized below.
\begin{itemize}
    \item \textbf{Protecting against Acoustic Eavesdropping.} In many cases, the audio channel can be used as a side-channel attack vector, to gain information exchanged using another channel, such as vibration. To protect against these events, dedicated sounds are introduced. These \emph{synthetic} sounds aim to pollute the audio communication channel, decorrelating it with the main communication channel and thus nullifying its eavesdropping potential. The schemes used to achieve this objective can be manifold, e.g., based on white noise, random sounds, or dedicated sounds mimicking the ones produced by the main communication channel, such as fake vibrations or keystrokes. 
    \item \textbf{Masking Sound Challenges.} The robustness of any audio-based defense scheme leveraging dedicated sounds emitted on purpose should be evaluated against the \acl{ICA} (ICA) attack. Among the wide possible number of solutions to decouple signals emitted by different sources, \ac{ICA} has been demonstrated to be the most effective one, and to be particularly successful when the emitting sources are sufficiently far each from the others. Therefore, masking sounds are suggested as an audio-based defense scheme in situations where the emitting sources are very close to each other, such as in the context of \aclp{IMD} or voice-controlled IoT devices. When a consistent separation exists between the sources of the audio signal, the designer of the solution should ensure that the attacker could not deploy a sufficient number of receivers to be able to perform subsequent analysis. 
    \item \textbf{Usability and Transparency of Audio-based Defense Schemes.} Any audio-based defense solution should deal with the resulting usability of the system. In the literature, the usability of the solution is often evaluated by looking at the requirement of additional components to the system, as well as to the explicit involvement of the user in the application of the defense strategy. Therefore, any valuable defense scheme should require neither any additional component nor the explicit involvement of the legitimate users. At the same time, possible distraction factors on the user's side should be reduced. For instance, when dealing with fake keystroke injection, despite the proposed defense scheme could be valuable and effective, the resulting distraction factors on the user could make the defense solution hardly usable.
\end{itemize}}

\section{Research Challenges and Future Research Directions}
\label{sec:challenges}

In the following, we identify the relevant research challenges related to each of the application domain analyzed in previous sections. We believe that the following discussion could be relevant for researchers actively working in the area, for inspiring new solutions and enhancing existing ones.\\

\textbf{Privacy-preserving Two-Factor Authentication}. The existing approaches, thoroughly described in Section \ref{sec:tfa}, have been demonstrated to be able to reach a very high level of usability, similar if not superior to existing token-based approaches. However, privacy concerns still represent a pressing issue. Indeed, leveraging pure ambient sounds exposes the whole scheme to biasing by malicious adversaries, as well as to co-located \ac{MITM} attacks. A successful solution to these issues leverages the generation of random sounds on the server, but it forces the authenticating device to deliver the local recordings to the server itself, possibly leaking information about the surrounding environment. Thus, privacy-preserving \ac{TFA} mechanisms based on short-range audio signals are still missing. Possible solutions to face these privacy issues could leverage innovative encryption techniques, such as \ac{HE} strategies, widely emerging in the last years~\cite{Acar2018}. Indeed, the \ac{HE} paradigm is a particular type of encryption, that can address the above privacy concerns by allowing a third actor to use encrypted data without the need to decrypt them. \\
\textcolor{black}{
In the context of audio-based TFA, the smartphone in possession of the user and the browser can first record the ambient sound, then encrypt the data recording using one of the many available HE techniques, and finally, deliver it to the server in the form of encrypted data. At the reception of the data, the server could compare the sounds recorded locally and remotely, operating over encrypted data, minimizing privacy concerns. These techniques have been already applied in various application contexts to similar privacy issues, such as in~\cite{Gomez2017} and~\cite{Erkin2012}.\\
We remark that, despite FHE strategies are normally associated to heavy computational requirements, few contributions in the literature already demonstrated that the main source of the computational overhead is not related to encryption operations, but only to operations over the encrypted data \cite{Veugen2015}. In the short-range audio TFA scenario, comparisons operations over encrypted data are executed on powerful servers, that can reduce the time needed to provide a decision by allocating more resources for this task.\\
Thus, the browser and the smartphone could deliver encrypted data to the server, without any risk for potential privacy issues. \\
Despite the promising directions, however, the first Fully Homomorphic Encryption (FHE) scheme has been revealed only in 2009, and still needs further improvements and refinements to be practical on nowadays computing platforms~\cite{Gentry2009fully}.\\\\
\textbf{Lightweight Solutions to MITM Attacks}. Despite the recent contribution in~\cite{Han2018} provided a viable solution to overcome co-located MITM attacks, further work is still necessary. Indeed, running a fingerprint evaluation process as the one provided by the authors in~\cite{Chen2017} could not be viable on more constrained devices, both because of the high processing capabilities required to provide a classification in a reasonable time, and because of the high number of audio messages to be exchanged. Alternative solutions could leverage an enhanced selection of features to fingerprint the audio environment in a specific location around a device, further minimizing the area around the target that could lead to a co-located MITM attack. Here, the challenge also lies in the minimal complexity of the fingerprinting process on devices such as commercial smartphones, not featuring the same processing and storage capabilities of server platforms.\\\\
\textbf{Non Invasive Audio-Based Pairing Schemes}. In the context of pairing, reducing the involvement of the human user is the most pressing issue. Indeed, removing the human from the process could also pave the way to less invasive solutions, not requiring audible sounds. The resulting stealthiness of the paring process could theoretically broaden the application scenarios where audio-based pairing schemes could be performed. This shift, however, cannot prescind from auxiliary means to verify that the pairing process is ongoing exactly between the intended devices. \\\\
\textbf{Facing Overshadowing Attacks in Constrained Devices}. Novel solutions to face inaudible overshadowing attacks are highly required. Despite the solution based on security signals proposed in~\cite{Hu_TDSC2018} is effective, it could be very energy and time demanding in specific contexts, i.e., involving Bluetooth and other battery-powered devices, where turning on the microphone or the speaker many times could be undesirable. In addition, requiring the sharing of secrets between the devices, such a solution is not practical when pairing devices previously unknown to each other. \\\\
\textbf{Acoustic Eavesdropping Risks Analysis}. Acoustic eavesdropping in the context of pairing deserves more attention. Indeed, even if the acoustic eavesdropping is a well-known threat, actually it does not seem a crucial one, given that the data delivered on the audio channel cannot be used to launch an attack. Indeed, it is worth noting that encryption techniques, as well as other security techniques, are very rarely applied to the audio signals, and their misuse could easily lead to severe threats. For instance, it is not clear if simple \emph{Replay Attacks}, performed by recording and playing back again the recorded audio signals, could lead to more severe threats when launched on audio-based pairing protocols.\\\\
\textbf{Overcoming the Hardware Heterogeneity}. The heterogeneity of microphones and speakers, as well as their different \emph{tolerance} to the noise in different application scenarios, take part in degrading the performance of audio-based pairing schemes when moved away from a suitable application scenario. For instance, as shown by the authors in~\cite{Fomichev2019}, the same Zero-Interaction audio-based solution report very different performance in indoor and outdoor scenarios. To overcome these limitations, researchers should design coefficient and comparison methods able to both remove inherent inaccuracies in the devices' hardware, and evaluate similarities and differences between recorded sounds in heterogeneous application domains.\\
In real deployments, the devices taking part in the pairing process could be very heterogeneous, being different not only in the number and the nature of embedded sensors, but also in the hardware chips providing sensing capabilities. These considerations are at the basis of several context-based pairing approaches, relying on physical properties that are invariant compared to the configuration of the pairing device. In this context, the microphones sensing the state of the audio channel could provide a different \emph{intensity} and \emph{timing} of a particular event. Thus, pairing approaches leveraging the audio channel should be able to remove time and intensity differences between pairing devices.\\\\
\textbf{Securing Voice Assistants}. Despite the relevant advances of the last years, securing voice assistants is still an issue. While actual solutions for authorizing the usage of voice assistants are referred to as \emph{Authentication} techniques, in practice they only provide a probabilistic assurance that who is issuing commands is effectively authorized. Limitations still exist, e.g., the approach in~\cite{Blue2018_asiaccs} has an uncertainty area of $\pm$15 degrees around the target. Thus, it is very effective in the proximity of the source, but it loses efficiency when the source is far. Similarly, the speaker fingerprinting technique applied in~\cite{Blue2018_wisec} may become inaccurate when in the background there is not silence, but other sounds. This is a typical situation in navigation assistants or voice assistants running on smartphones.  At the same time, biometric recognition techniques to identify the voice of a (set of) legitimate owner(s) have still received little attention from the community, and needs to be further investigated to prove its effectiveness. This is especially true when they are applied on constrained devices, such as IoT audio-controlled devices. Indeed, efficient and effective techniques are needed, possibly coupling already available solutions in an engineered fashion. \\\\
\textbf{Sound Masking Effectiveness Evaluation}. Currently, many audio-based defense techniques are based on \emph{sound masking} strategies, described in Section \ref{sec:masking}. However, some contributions such as~\cite{Zhang2014} already provided evidence that this technique can fail to provide the desired masking properties if the sources of the audio signal, i.e., the genuine sound and the masking sound, are located sufficiently far from each other. This limitation seems to be a big issue in specific contexts. For instance, when protecting sounds emitted by keys on a keyboard, the speaker of the laptop and the keyboard could be sufficiently far from each other to allow an attacker to identify the location of the device emitting the particular sound. At the same time, the effectiveness of injecting fake sounds is still not widely assessed in practical conditions. Indeed, the effectiveness of these techniques should be carefully evaluated in the presence of attackers able to recognize and to tell apart sounds emitted by a speaker and mechanical sounds, recalling the technique used in~\cite{Chen2017}.\\\\
\textbf{Proximity Detection Schemes Improvement}. Another defense tool leveraging 
audio signals is proximity detection. This approach refers to the similarity of ambient audio to assess co-presence between communicating devices. As shown in a recent contribution~\cite{Shrestha2019}, it is feasible for a powerful attacker to manipulate the context and bypass the proximity detection evaluation by biasing the environment, e.g., inserting sounds or predicting the environment. Thus, audio-based proximity detection could be probably more useful if used in combination with other co-presence detection methods, in a way to force the attackers to realize more expensive and powerful attacks. Indeed, proximity detection could not provide a definitive assurance that the attacker cannot overcome the defense strategies. Thus, contributions are still needed in this area.\\\\
\textbf{Attacks and Defenses via Directional Microphones}. The large majority of attacks available in the literature and discussed in the previous section focused on omnidirectional microphones. Indeed, as per their definition, they can pick up sounds from almost every direction with the same performances. While this can be a useful feature in some scenarios, there are situations where an enhanced sensitivity towards particular sounds is desired. In this context, directional microphones can provide a meaningful solution. Indeed, a directional microphone is more sensitive to picking up sounds in certain directions rather than others. At the same time, specific types of directional microphones increase the reception range along a specific direction, as they can detect sounds emitted at a higher distance than regular omnidirectional ones~\cite{Nagaraja2018},~\cite{Schwartz2018}. Indeed, these interesting features make them an attractive choice both for attacks and defense. On the one hand, attackers could increase the minimum required distance to launch an attack against a target device. On the other hand, directional microphones can implicitly reject attackers located outside the main reception lobe. 
Despite their evident advantages, the use of directional microphones for both attacks and defense is still not widespread yet. 
Hence, it represents a promising and interesting research direction to improve state-of-the-art results and to come up with innovative security tools. \\\\
\textbf{Advanced Security Schemes via 3D Audio}.  Thanks to the very high pace of technological innovations, the first 3D Audio systems are hitting the market, with the promise to significantly enhance the sound experience of moving users and devices~\cite{Johansson2019}. 3D audio recordings are created by placing a large number of microphones in the area. In this way, the sound of the scene is recorded at the same time from different positions, allowing to re-create the real sound experience of a moving entity. From the security perspective, 3D audio schemes could have a disrupting potential, both from the attack and from the defense side. To date, there are no studies in the literature that investigate this 
venue, that is probably the most novel and promising one in the short-range audio-based security context.}

\section{Conclusions}
\label{sec:conclusions}

In this paper, we have provided a thorough survey of mechanisms,  applications, use-cases, and research challenges for short-range audio channels security.
We showed that, thanks to enhanced usability features and low deployment costs, techniques based on short-range audio channels can be used as a means to achieve innovative and effective security services, such as Two-Factor Authentication, pairing, and device authorization schemes, to name a few. 
These properties are enforced by leveraging specific physical-layer features of the audio channels, such as distance bounding and physical proximity detection of devices sharing the same audio context. We have also shown that, if not integrated correctly, such methodologies could be subject to a variety of attacks. 
Thus,  research is needed to improve defense solutions based on short-range audio signals. 

Finally, we have also highlighted upcoming research challenges.
The exposed challenges show that the development of audio-based solutions is still an exciting research area, and that it can be inspiring for researchers, industry, and start-ups, striving for innovative, non-invasive, and computationally lightweight means to enforce systems security.

\section*{Acknowledgements}
The authors would like to thank the anonymous reviewers for their comments and suggestions, that helped improving the quality of the manuscript.\\*
This publication was partially supported by awards NPRP11S-0109-180242, UREP23-065-1-014, and NPRP X-063-1- 014 from the QNRF-Qatar National Research Fund, a member of The Qatar Foundation. The information and views set out in this publication are those of the authors and do not necessarily reflect the official opinion of the QNRF.

\bibliographystyle{IEEEtran}
\bibliography{audio}
\begin{IEEEbiography}[{\includegraphics[width=1in,height=1.15in,clip,keepaspectratio]{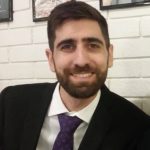}}]{Maurantonio Caprolu}
Maurantonio Caprolu is currently a PhD student in Computer Science and Engineering at Hamad Bin Khalifa University. He received both his Bachelor and his Master's Degree in Computer Science at Sapienza, University of Rome, Italy, on topics strictly related to applied Security and Privacy. His major research interests include security issues in Blockchain-based systems, Edge/Fog architecture and Software Defined Networking, Computer Forensics, Information Warfare. 
\end{IEEEbiography}
\begin{IEEEbiography}[{\includegraphics[width=1in,height=1.15in,clip,keepaspectratio]{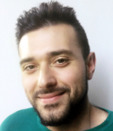}}]{Savio Sciancalepore}
Savio Sciancalepore is currently Post Doc at HBKU-CSE-ICT, Doha, Qatar. He obtained his Master's degree in Telecommunications Engineering in 2013 and the PhD in Electric and Information Engineering in 2017, both from the Politecnico di Bari, Italy. He received the prestigious award from the ERCIM Security, Trust, and Management (STM) Working Group for the best Ph.D. Thesis in Information and Network Security in 2018. His major research interests include security issues in Internet of Things (IoT) and Cyber-Physical Systems.
\end{IEEEbiography}
\begin{IEEEbiography}[{\includegraphics[width=1in,height=1.15in,clip,keepaspectratio]{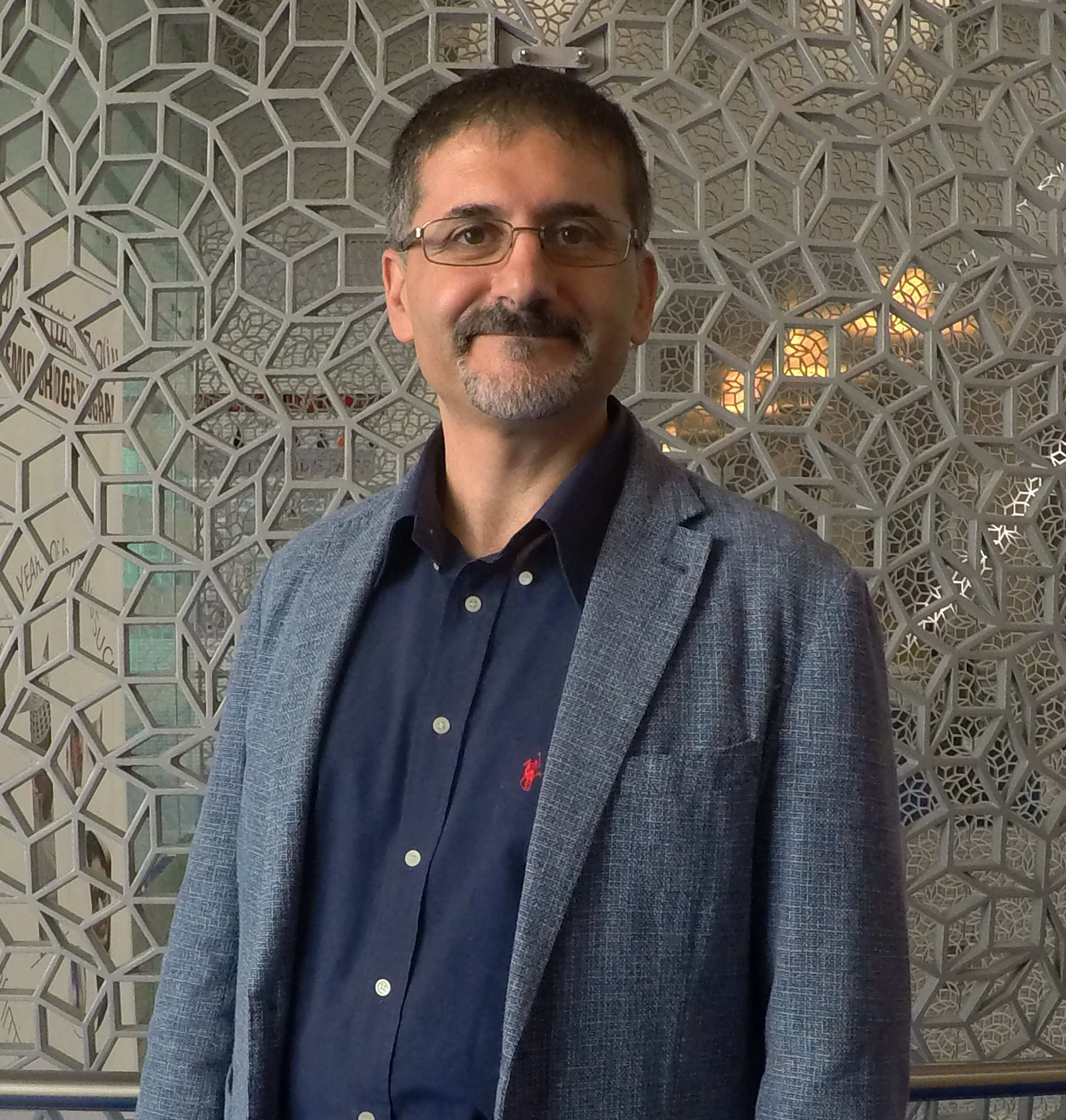}}]{Roberto Di Pietro,} ACM Distinguished Scientist, is Full Professor in Cybersecurity at HBKU-CSE. Previously, he was in the capacity of Global Head Security Research at Nokia Bell Labs, and Associate Professor (with tenure) of Computer Science at University of Padova, Italy. He has been working in the security field for 23+ years, leading both technology-oriented and research-focused teams in the private sector, government, and academia (MoD, United Nations HQ, EUROJUST, IAEA, WIPO). His main research interests include security and privacy for wired and wireless distributed systems (e.g. Blockchain technology, Cloud, IoT, OSNs), virtualization security, applied cryptography, computer forensics, and data science. 
Other than being involved in M\&A of start-up---and having founded one (exited)---, he has been producing 220+ scientific papers and patents over the cited topics, has co-authored two books, edited one, and contributed to a few others. 
He is serving as an AE for  ComCom, ComNet, Journal of Computer Security, and other Intl. journals. In 2011-2012 he was awarded a Chair of Excellence from University Carlos III, Madrid. Google says that he has been receiving 8300+ citations, with an h-index=44 and an i-index=120.
\end{IEEEbiography}

\end{document}